\documentclass[11pt,a4paper]{JHEP}
\usepackage{amsmath}
\usepackage{amsfonts}
\usepackage{epsfig}  
\usepackage{feynmf}
\usepackage{amssymb}
\def\ln{{\rm ln}}

\def\D{{\cal D}}
\def\A{{\cal A}}
\def\M{{\cal M}}
\def\G{{\cal G}}
\def\I{{\cal I}}
\def\J{{\cal J}}
\def\K{{\cal K}}

\def\a{\alpha}
\def\r{\rho}
\def\zb{\bar z}
\def\rb{\bar\rho}
\def\sin {{\rm sin}}
\def\cos{{\rm cos}}
\def\d{\partial}

\def\tr{{\rm tr}}
\def\Tr{{\rm Tr}}
\def \nc {noncommutative }
\newcommand\0{\nonumber}
\newcommand\ee{\end{eqnarray}}	 	
\newcommand\be{\begin{eqnarray}}
\newcommand\ba{\begin{array}}			
\newcommand\ea{\end{array}}

\preprint{SISSA/45/01/EP\\hep-th/0105307\\new version}

\title{Non--orientable string one--loop corrections in the presence 
 of a B field}

\author{ L.\ Bonora, M.\ Salizzoni\\
International School for Advanced Studies (SISSA/ISAS)\\
Via Beirut 2--4, 34014 Trieste, Italy, and INFN, Sezione di Trieste\\
E-mail:   \email{bonora@sissa.it}, \email{sali@sissa.it}}
 
\abstract{We discuss the problem of \nc $SO(N)$ gauge field theories
from the string one--loop point of view. To this end
we propose an expression for the string propagator on the 
boundary of the M\"{o}bius strip in the presence of a constant 
$B$ field. We discuss in detail the problems related to
its derivation.
Then we use it to compute the one--loop corrections to two--, 
three-- and four--gluon amplitudes in an open string theory with 
orthogonal Chan--Paton factors.  We show that these corrections 
in the field theory limit in 4D are compatible with the one--loop 
corrections of a renormalizable \nc $SO(N)$ gauge field theory.}

\begin{document}
\begin{fmffile}{fmfncnor}

\section{Introduction}

An interesting problem that has been raised in connection with 
the recent attention on noncommutative field theories as effective 
field theories of open strings attached to D--branes in the presence 
of a constant B field, is the existence of noncommutative 
gauge theories with gauge transformations valued in a Lie subalgebra 
of $u(N)$. There 
are several reasons why the existence of at least
some of them is expected and desirable. From an abstract point of 
view, \cite{connes}, there should not be any obstruction to 
constructing a noncommutative gauge field theory with any Lie 
algebra (even though this may not imply that these theories are 
effective field theories of 
the strings). On the other hand we know that \nc field theories 
retain certain features of string theory better than ordinary theories 
do, [2-31]. We have in mind here the ultraviolet convergence
properties of \nc theories but, even more, the possibility of having
soliton solutions in situations where ordinary theories are unfit to 
support them, \cite{harvey}. This is particularly important in 
connection with tachyon condensation. In this regard, another important 
property is the possibility of embedding the Moyal product into the 
star product of open string field  theory in a factorized way, 
\cite{witten}.
It would be rather disappointing  if such remarkable properties could 
not be extended, for example, to string  theories or string field 
theories with orthogonal Chan-Paton factors. 
 
Recently there have been a few attempts at defining and studying \nc 
versions of gauge field theories with orthogonal and symplectic, 
\cite{BSST,BarsShahin}, or even more general Lie algebras, \cite{MSSW}.
These \nc theories have been defined at the semiclassical (tree) level. 
As soon as one tries to go beyond the tree level one has to face an 
unexpected result: in four dimensions they look (at least naively) 
nonrenormalizable. One is tempted to dismiss this fact as a 
non--problem. After all, these are effective field theories, which are 
nonlocal as ordinary theories. However the right question we should ask 
is whether this corresponds to some feature (perhaps ill--definiteness) 
of the string theory the gauge  field theory is supposed to represent 
in the low energy limit. To know  the answer we have to study 
the one--loop corrections of the relevant string theory. This is what 
we want to do in this paper for an unoriented open string theory 
with orthogonal Chan--Paton factors in the presence of a background 
$B$ field. We would like to specify from the very beginning
that our approach is not unproblematic, and we will list
below the aspects of our treatment that may appear controversial.

In order to compute one--loop corrections one needs the 
string Green functions on the relevant
world--sheets, which are the annulus and the M\"{o}bius strip. 
While the Green function for the former case in the presence of a 
$B$ field is well--known, the latter case has not been studied yet.
For our purposes we need the propagator on the boundary of the strip, 
but the presence of the $B$ field requires the knowledge of the 
propagator on the whole M\"{o}bius strip in order for us to be able to 
take an unambiguous limit for the propagator on the boundary.
Extending the propagator outside the boundary of the M\"{o}bius strip
is a non--trivial operation, even in the ordinary case ($B=0$).
For this reason we devote section 2 to the construction of the string 
propagator on the M\"{o}bius strip in the ordinary case, a subject which 
does not seem to have been carefully analyzed in the literature. 
In section 3 we use this propagator in order to compute one--loop 
2--, 3-- and 4--point gluon amplitudes and their field theory limit 
in the absence of a $B$ field. 

Subsequently we turn on a constant $B$ field. In section 4 we compute
the string propagator on the M\"{o}bius strip, we discuss the problems
raised by this calculation and finally we find the expression of
the propagator on the boundary, which is what we actually need since 
we intend to compute amplitudes of vertex operators inserted 
at the boundary of the world--sheet.
With this tool it is then elementary,
in section 5, to extend the one--loop results of section 3 to the case
of a constant $B$ field. On the basis of the results obtained in 
section 5 we would be led to conclude that the string one--loop 
corrections entails that {\it the \nc limiting field theory corrections 
(in 4D) are those of a one--loop renormalizable field theory with 
the same renormalization
constants as the corresponding ordinary $SO(N)$ gauge field theory}.
However, apparently, this does not correspond to what one gets
from one--loop corrections in the corresponding \nc $SO(N)$
gauge field theory. In section 6 this issue is discussed at length.

Finally, as promised above, we would like to list the most 
problematic aspects of our paper. These are: (i) the possibility of
a non--vanishing $B$ field in the string context considered here,
(ii) the continuation to the bulk of the M\"{o}bius
strip of our string propagator, (iii) the one--loop 
non--renormalizabity of the \nc $SO(N)$ gauge theories. In the
course of the paper we argue that all these aspects may not be
unsurmountable obstacles. In any case we do not see decisive 
arguments in favor of the contrary. Altogether we believe that
the approach presented here, although not accompanied by  
uncontroversial arguments, represents nevertheless a concrete 
possibility.

\section{The string propagator on a non-orientable world-sheet. 
Case $B=0$}

One--loop contributions in open unoriented string theory come from
the annulus and the M\"{o}bius strip world--sheet. Henceforth
for conciseness we denote a M\"{o}bius band by ${\cal M}$ and an 
annulus by $\A$. As for the parametrization of the latter we will 
use the notation of \cite{GSW}. The annulus will be represented 
either in the $z$--plane or in the $\rho$--plane. In the first case 
the annulus is represented in the most obvious way as the region 
$q\leq |z|\leq 1$, where $q$ is the modulus. In the $\rho$-plane 
the annulus will be identified with the region 
$w\leq |\rho|\leq 1$ of the lower half plane with the lower
and upper semicircle identified in such a way as to preserve 
the orientation of the surface (the two semicircles are `parallel'). 
The map between the two representations is given by:
\be
z = e^{2\pi i\frac {\ln \rho}{\ln w}}, \quad \quad \ln q = 
\frac {2\pi^2} {\ln w}\0
\ee
Alternatively the modulus is represented by the imaginary number 
$\tilde\tau$ defined by:
\be
q= e^{i\pi \tilde\tau},
\quad \quad {\tilde\tau} = - \frac {2\pi i}{\ln w}\0 
\ee
It is convenient to perform the modular transformation 
$\tilde \tau\to -1/{\tilde\tau}$. After this operation, 
following \cite{VMLRM}, we will
parametrize the above variables as follows
\be
w=e^{-2\tau}, \quad\quad \rho=e^{-2\nu}\label{tau}
\ee
where $\tau= -i \pi \tilde \tau$.

The representation of the M\"{o}bius band is the same except that
the upper semicircle in the $\rho$--plane is identified with the lower
one in an antiparallel way (see figure 1). The field theory limit 
corresponds to an infinitely thin annulus or band, i.e.
$q \to 1$, which corresponds to $w \to 0$ or $\tau\to\infty$.

Our purpose in this paper is to compute amplitudes involving several 
gluon vertices inserted at the boundary of the annulus $\A$ or of 
the M\"{o}bius strip $\M$. To this end
we need to know the string propagator on both surfaces.
The string propagator in the annulus, in the presence of a $B$ field,
was calculated long ago in \cite{ACNY} and elaborated on in \cite{BCR}.

As for the string propagator on the boundary of $\M$, in the absence of 
$B$, it can be found, for example, in \cite{GSW}. However, as explained
in the introduction, when in presence of a $B$ field one needs to know
the propagator in the bulk of $\M$ in order to be able to take the 
correct limit to the boundary. In view of this it is a good propedeutical
exercise to find the string propagator on the bulk of $\M$ without $B$ 
field. This exercise does not seem to have been done previously in 
the literature.

\begin{figure}[ht]
\begin{center}
{\scalebox{1}{\includegraphics{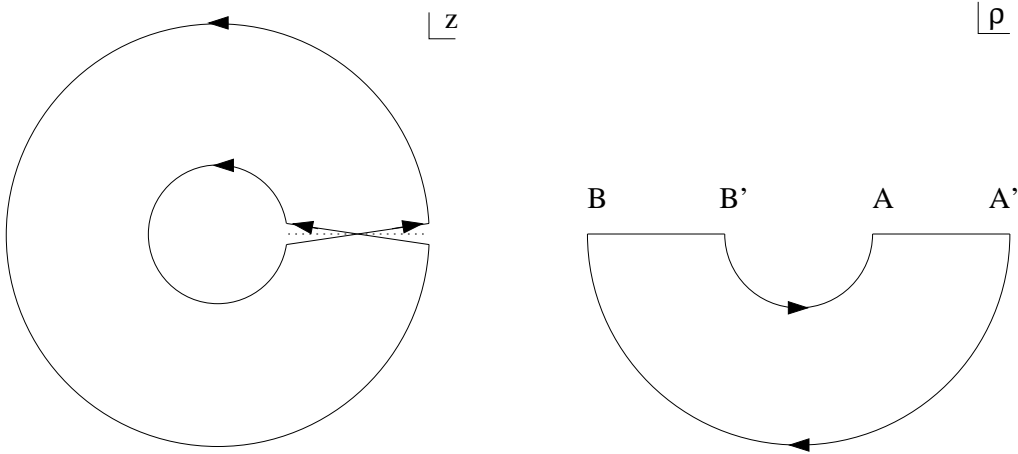}}}
\end{center}
\caption{The representation of the M\"{o}bius band in the $z$ and $\rho$ planes}
\end{figure}         

\subsection{The Green function method}

We start from the sigma model action of open strings attached to a D--brane
 \be
\frac{1}{4\pi \alpha'}\int_\Sigma d^2x \left(\sqrt{h} 
h^{\alpha\beta}\d_\alpha X^i\d_\beta X^j g_{ij}
\right)\label{action0}
\ee
where $g_{ij}$ is the (closed) string metric.
Our problem is to find, on the surface $\Sigma$ of interest
(either $\A$ or $\M$), a solution (the Neumann function) of the equation
\be
\nabla^2 \G^{ij}(x,x') = 2\pi \alpha' g^{ij} \delta (x-x')
\label{green}
\ee
satisfying the boundary conditions
\be
\left. \d_\perp \G^{ij}(x,x') \right \vert_{\d\Sigma}=0
 \label{bc}
\ee
In these equations $x$ stands for either $z$ or $\rho$ (see Appendix A
for some auxiliary formulas).

However with the (\ref{bc}) boundary conditions it would be impossible 
to satisfy Gauss's theorem. Therefore we modify them so that
in $\rho, \rb$ coordinates the above equations become
\be
4\d_\rho\d_{\bar \rho} \G^{ij}(\rho,\rho')= 
2\pi \alpha' g^{ij} \delta(\rho-\rho')\label{greenrho}
\ee
and
\be
\d_\theta
\left. \G^{ij}(\rho,\rho')\right\vert_{\d\Sigma}= K g^{ij}\label{bcrho}
\ee
The constant $K$ will be determined below.
The boundary $\d\Sigma$ corresponds to real $\rho$ with 
$w\leq |\rho|\leq 1$.

Following the notation of \cite{ACNY,BCR} we write the solution for 
the M\"{o}bius strip as follows:
\be
\frac {1}{\alpha'}\G_\M^{ij}(\rho,\rho') = g^{ij}\left(\I^\M(\rho,\rho') 
+\J^\M(\rho,\rho')+f^\M(\rho,\rho')\right)\label{GM}
 \ee
where
\be
\I_\M(\rho,\rho') &=& \ln\left(\frac{|\tilde\tau|}{2}\right) + 
\frac{(\ln\frac{\r} {\r'})^2+(\ln\frac{\rb} {\rb'})^2}{4\,\ln w}+
\ln \left\vert\sqrt{\frac{\r}{\r'}}-\sqrt{\frac{\r'}{\r}}\right\vert\0\\
&&\quad\quad\quad+ \ln \prod_{n=1}^{\infty}\frac{\left\vert 1-(-w)^n\frac{\r}{\r'}\right\vert \cdot
\left\vert1-(-w)^n\frac{\r'}{\r}\right\vert}{(1-(-w)^n)^2}\label{IM}\\
\J_\M(\rho,\rho') &=& \ln\left(\frac{|\tilde\tau|}{2}\right)
+\frac{(\ln\frac{\r} {\rb'})^2+
 (\ln\frac{\rb} {\r'})^2}{4\,\ln w}+
\ln \left\vert\sqrt{\frac{\r}{\rb'}}-\sqrt{\frac{\rb'}{\r}}\right\vert\0\\
&& \quad\quad\quad +\ln \prod_{n=1}^{\infty}\frac{\left\vert 1-(-w)^n\frac{\r}{\rb'}\right\vert\cdot
\left\vert 1-(-w)^n\frac{\rb'}{\r}\right\vert}{(1-(-w)^n)^2}\label{JM}\\
f_\M(\rho,\rho')&=& -\frac {i\pi}{2\ln w}\, \ln\, \frac {\r\r'}{\rb \rb'},
\label{fM}
\ee
There is a subtlety in the above definition: the log square terms must 
be understood as
\be
\Big(\ln \frac{\r}{\r'}\Big)^2 = \frac 14 \Big(\ln \big(\frac{\r}{\r'}\big)^2
\Big)^2 \0
\ee
and so on.

Notice that $\G_\M^{ij}(\rho,\rho') = \G_\M^{ji}(\rho',\rho)$. It is now
quite a standard matter to verify that eqs.(\ref{greenrho}) and 
(\ref{bcrho}) are satisfied, with $K$ being
\be
K= \frac{2\pi}{\ln\,w}\label{K}
\ee
It is also easy to verify that the 
continuity conditions on the boundary of the M\"{o}bius band 
are satisfied:
\be
\G_\M^{ij}(1,\r') = \G_\M^{ij}(-w,\r'), \quad \quad \G_\M^{ij}(-1,\r') =
\G_\M^{ij}(w,\r'),\quad \quad\forall \, \rho'\0
\ee
However one should notice that the propagator is not everywhere
continuous. In fact, while it is easy to verify that 
\be
(\I_\M + \J_\M)(-w\rb,\r') = (\I_\M+\J_\M)(\r,\r')\label{I+J}
\ee
so that the combination $(\I_\M + \J_\M)$ satisfies everywhere the 
periodicity conditions for the M\"{o}bius strip, a similar identity is 
not satisfied by $f_M$, for which we have
\be
f_\M(e^{-i\theta},\r')- f_\M(-we^{i\theta},\r')=\ln(e^{-4i\theta})
\label{discf}
\ee
so that there is a line of discontinuity that does not permit an exact
matching along the arcs $A'B$ and $B'A$. One simple solution would be
to drop $f_\M$ in the definition (\ref{GM}). But in this case we 
would not satisfy Gauss's theorem, see below. On the other hand $f_\M$ does not
contribute to the propagator on the boundary, thus we prefer to 
keep $f_\M$ in the definition (\ref{GM}) (for further comments on this
point see below and section 4).  

It is now time to discuss Gauss's theorem, i.e. the integrated version
of (\ref{greenrho}), which says that the integral of $\d_\perp \G_\M$
along the boundary $\d\M$ equals the integral of the RHS of 
(\ref{greenrho}). The latter equals ${2\pi} \alpha'$. As for the
former one has to integrate over the boundary $AA'$ and $B'B$ of $\M$,
but, due to the above mentioned discontinuity, also along the arcs
$A'B$ and $B'A$. Now the normal derivative of $G_\M$ along 
the arcs $A'B$ and $B'A$ vanishes as well as the normal derivative 
of $\I_\M+ \J_\M$ along $AA'$ and $B'B$. The only non--vanishing 
contribution comes from the normal derivative of $f_\M$ along
the boundary of $\M$. This is given by
(see Appendix A for notation)
\be
\int_A^{A'} dl \, \d_\perp f + \int_{B'}^B dl\, \d_\perp f = 2\pi\label{gauss}
\ee
and we have used the fact that $(\r \d_\r - \rb \d_{\rb})\ln \frac {\r}
{\rb}=2$. Therefore Gauss's theorem is verified. 

However so far we have been somewhat cavalier in integrating over $\M$.
The point is that the M\"{o}bius strip is a nonorientable
surface and integration theory on nonorientable surfaces takes on
a peculiar twist. The reason is that on nonorientable manifolds 
only {\it densities} can be integrated, see \cite{botu}. A density 
is an expression that, under a coordinate change, gets 
multiplied by the inverse modulus of the Jacobian of the partial 
derivatives (not just by the inverse Jacobian). When we integrate
both sides of (\ref{greenrho}) we are precisely integrating two 
densities. However, for this reason, the measure on $\M$ cannot induce 
an oriented measure on the boundary. The consequence is that the 
relative signs on the two sides of Gauss's theorem remains undetermined.
This question is purely technical and can be settled by means
of a physical analog: consider an electrostatic 
analog in which $\G$ is proportional to the electrostatic potential 
generated by a charge placed in $\r'$. The integral of the normal
derivative (electric field) along the boundary equals the total charge.
Therefore we know how to fix the relative sign in Gauss's theorem.
It has also to be remarked that the normal derivative across the line
$A'B$ and $B'A$ is continuous, nothwistanding the discontinuity 
mentioned above. 

In the following we will need the propagator on the boundary of $\M$.
This is obtained by taking the limit for $\rho$ and
$\rho'$ approaching the real axis: 
$\G_{\M}^{ij} \rightarrow G_{\M}^{ij}$. 
We get
 \be
G_{\M}^{ij}(\rho,\rho')= 2\alpha'\hat g^{ij} G_\M(\rho,\rho') 
 \label{boundprop}
\ee 
where $G_\M(\rho,\rho')$ is given by
\be
G^+_\M(\rho,\rho')&=& \ln\left(- \frac{\pi}{\ln w}\right) + 
\frac{\big(\ln\frac{\r} {\r'}\big)^2}{2\,\ln w}+
\ln \left\vert\sqrt{\frac{\r}{\r'}}-\sqrt{\frac{\r'}{\r}}\right\vert\0\\
&&\quad\quad+ \ln \prod_{n=1}^{\infty}\frac{\left(1-(-w)^n\frac{\r}{\r'}\right)
\left(1-(-w)^n\frac{\r'}{\r}\right)}{(1-(-w)^n)^2},
\quad {\rm if}\quad \rho\rho'>0\label{I+M}\\
 G^-_\M(\rho,\rho')&=& \ln\left(- \frac{\pi}{\ln w}\right) + 
\frac{\big(\ln\left\vert\frac{\r} {\r'}\right\vert\big)^2}{2\,\ln w}+
\ln \left\vert\sqrt{\left\vert\frac{\r}{\r'}\right\vert}+
\sqrt{\left\vert\frac{\r'}{\r}\right\vert}\right\vert\0\\
&&\quad\quad+ \ln \prod_{n=1}^{\infty}\frac{\left(1+(-w)^n\left\vert\frac{\r}{\r'}
\right\vert\right)
\left(1+(-w)^n\left\vert\frac{\r'}{\r}\right\vert\right)}{(1-(-w)^n)^2},
\quad {\rm if}\quad \rho\rho'<0\label{I-M}
\ee
This is the propagator we will use for our calculations in the following 
section. 

Finally we notice that by replacing $(-w)^n$ with $w^n$ in (\ref{GM})
we get the Green function for the annulus, from which one
can extract the planar and nonplanar propagators.

\section{Field theory limit of gluon amplitudes without $B$ field}

We wish to calculate string theory amplitudes and to extract from 
them information concerning the low energy effective field theory. 
In particular 
we are interested in the renormalization properties (in 4D) of the 
latter. For this reason in this section we intend to compute two--, 
three-- and four--gluon one--loop
 amplitudes from string theory with $SO(N)$ CP 
factors and evaluate their field theory 
 limit, more specifically the
UV divergent contributions of the various  amplitudes, in 
order to compare them with the field theory ones. While this has been 
done in detail in theories with unitary CP factors, 
\cite{VMLRM}, to our best knowledge nothing similar has been done for 
theories with orthogonal or symplectic CP factors. Therefore working 
out the field theory limit in the latter case without $B$ field is a 
necessary preparation to the next section and a calculation interesting 
in itself. The novelty in this case is that, beside the annulus 
amplitudes, one has to consider also the M\"{o}bius strip ones.

The method we adopt here was developed over the years by 
several people, see \cite{VMLRM} and references therein. It 
is based on calculations carried out in the framework of the bosonic 
string theory. Indeed it is enough to embed the gauge field theory we 
want to regularize in the bosonic string theory. It it is not even 
necessary that the string theory be critical. As a regulator of a field 
theory a bosonic string theory in generic dimensions
will do. For these and other considerations on the method used here, 
we refer to \cite{VMLRM}). 

We start by writing down the tree level gluon amplitudes with
CP factors belonging to the Lie algebra $SO(N)$ at the lowest order 
in $\alpha'$.
\be
A^{(0)}(p_1,p_2) &=& 2i\, \tr (t^{a_1}t^{a_2})\, 
\epsilon_1\cdot\epsilon_2 p_1\cdot p_2
\label{2p0}\\
A^{(0)}(p_1,p_2,p_3) &=&  4 g_D\tr (t^{a_1}t^{a_2}t^{a_3}) 
\left(\epsilon_1\cdot\epsilon_2 p_2\cdot \epsilon_3
+ \epsilon_2\cdot\epsilon_3 p_3\cdot \epsilon_1
+\epsilon_3\cdot\epsilon_1 p_1\cdot \epsilon_2\right)\label{3p0}\\
A^{(0)}(p_1,p_2,p_3,p_4) &=& 4ig_D^2 \tr (t^{a_1}t^{a_2}t^{a_3}t^{a_4})  
\left(\epsilon_1\cdot\epsilon_2 \epsilon_3\cdot \epsilon_4  
\frac{p_1\cdot p_3}{p_1\cdot p_2}\right.\0\\ 
&&~~~~~~~~~~~~~~~~~~~~~~~~~~~\left. +  
\epsilon_1\cdot\epsilon_3 \epsilon_2\cdot \epsilon_4  
+\epsilon_1\cdot\epsilon_4 \epsilon_2\cdot \epsilon_3  
\frac{p_1\cdot p_3}{p_2\cdot p_3} 
\right) \label{4p0}  
 \ee
To give a meaning to eq.(\ref{2p0}) it is useful to introduce a small
mass for the gluon: $p_i^2=m^2$ (which is anyhow necessary as an IR 
cutoff, although we will not need it explicitly in the following).
The above amplitudes have been normalized in such a way as to coincide 
with the corresponding tree level amplitudes in field theory.
In particular, $g_D$ is the D--dimensional gauge coupling, the $t^a$'s 
are the  generators
of $SO(N)$ in the fundamental representation, the $\epsilon_i$'s are 
gluon polarizations and $p\cdot q= p_i\hat g^{ij}q_j$. 
Later on we will use the above formulas for $D=4$. In that case
$g_D=g_4\equiv g$. We recall that (\ref{4p0}) contains, in field
theory terms, also one--particle reducible contributions.

We write down now general form of the one--loop amplitudes (which, for 
later reference, is valid in general, also when a $B$ field is switched 
on):
\be
A^{(1)}(p_1,\ldots,p_M) & = & \frac 12\,\chi_M\,f_N^{a_1,a_2,...,a_M}\,
\frac{ g_D^{M}}{(4\pi)^{\frac D2}}\,(2\alpha')^{-\frac{D}{2}}
\int\prod_{r=2}^Md\nu_r 
d\tau e^{2\tau}\tau^{-\frac D2}\label{1lampl} \0 \\
& \times & \prod_{n=1}^\infty
\left(1-\eta_ne^{-2n\tau}\right)^{2-D}
\exp\left[\sum_{r<s}p_r G(\nu_{rs})p_s\right] \0 \\
& \times & \exp \left[\sum_{r\neq s}\left( p_s \d_r G(\nu_{sr})
\epsilon_r
+ \frac 12 \epsilon_r \d_r\d_s G(\nu_{sr})\epsilon_s
\right)\right]_{\rm m.l.}
\label{Aoneloop}
 \0
\ee
where $\chi_M = i\,(1)$ for $M$ even (odd).
$f_N^{a_1,a_2,...,a_M}$ is the group 
theory factor. It equals $N \tr(t^{a_1}\ldots t^{a_N})$ in the   
annulus case for planar amplitudes and  $\tr(t^{a_1}\ldots t^{a_N})$  
in the M\"{o}bius strip case. Moreover  $pGq$ stands for $p_iG^{ij}q_j$,
$\nu_{rs}=\nu_r-\nu_s$ and $\d_r = \frac {\d}{\d\nu_r}$. 
The factor $\eta_n=1$ in the orientable case, $=(-1)^n$ in the
non-orientable case.The suffix ${\rm m.l.}$ stands for 
multilinear, meaning
that in the series expansion of the exponential
we keep only the terms that
are linear in each polarization. The
propagator $G$ is either the annulus or
the M\"{o}bius strip propagator, and
the integrals over the $\nu$ variables
are
evaluated in the appropriate
regions of integration (moduli space).
 
The constants in front of the tree and one--loop amplitudes have been 
defined in such a way as to agree in the zero slope limit with the 
corresponding field theory results.

The strategy now consists in replacing in
eq.(\ref{1lampl}) the appropriate
 propagators and singling out the regions of
the moduli space which give rise
 to divergent contributions in the $\a'\to 0$
limit. This will be done explicitly
 below for the M\"{o}bius amplitudes. As for
the annulus amplitudes, since their 
 evaluation does not depend on the CP
factors, we can borrow for them the analysis
 already carried out in
\cite{VMLRM} and \cite{BCR,CRS} in the case of unitary CP 
 factors. These
amplitudes split in general into planar and non--planar contributions. 
 As
for the latter we can rely on the results of \cite{BCR}, which, as expected,
tells us that they do not give rise to UV divergences in the field theory 
limit\footnote{A more careful statement is needed when a constant $B$ field 
is
present because of the UV/IR mixing. However in this paper we will not 
deal
with this problem.}. The planar amplitudes do give rise to divergent
contributions  
 in the field theory limit. They have been analyzed in detail
in \cite{VMLRM}: assuming
 dimensional regularization in the $\a'\to 0$ limit,
they reproduce exactly the
 results obtained in field theory with the
background field method \cite{abbot}.
 More precisely, one can single out the
divergent part that corresponds in field theory
 to one--particle irreducible
diagrams. The result can be written 
\be
 \left. A^{(1)}(p_1,...)\right|_{\rm div}=
-\frac N2\, \frac{g^2}{(4\pi)^2} \,\frac {11}{3} 
 \frac
{1}{\epsilon}A^{(0)}(p_1,...)\label{A1A0}
 \ee
for two--, three-- and four--point functions, with $\epsilon = 2-D/2$. 
Throughout the paper the 
label $div$ stands for irreducible divergent part, in the sense that
in field theory these divergences correspond to one-particle irreducible
diagrams. It is also possible to extract from string theory the
one--loop one--particle reducible contributions but here we will not be
concerned with them.
In the remaining part of this section we will show how to
extract relations similar to
 (\ref{A1A0}) for the M\"{o}bius amplitudes.
Following \cite{VMLRM}, we will use two 
 different methods.
Since these two methods
have already been carefully  spelt out in \cite{VMLRM} for the annulus
amplitudes, we skip many details and focus on the peculiarities introduced by
a non--orientable world--sheet.  

\subsection{M\"{o}bius amplitudes: first method}

This method is based on the `doubling trick', \cite{GSW}. One can show that
a large amount of information contained in a M\"{o}bius amplitude is
captured by doubling the integration region. Let us start from the propagator along the boundary of $\M$, written as follows:
\be
   G_{\M}(\rho, \rho') = \ln \left[ \frac{1-c}{\sqrt{c}}
                         \exp \left(\frac{\ln^2 c}{2\ln w}\right)
                         \prod^{\infty}_{n=1}     
\frac{(1-(-w)^n c)(1-(-w)^n/c)}{(1-(-w)^n)^2}\right],\0
\ee
 where $c=\rho/\rho'$. This coincides with (\ref{I+M}) provided
$c \leq 1$. Following \cite{GSW}, it can be recast in the form:
\be
G_{\M}(\nu-\nu') = \ln \left[ -\frac{4\pi}{\ln q}\, \sin\left( \frac{\pi(\nu-\nu')}{2}\right)
\prod^{\infty}_{n=1}\frac{1-2\,(-\sqrt{q})^n \,\cos(\pi(\nu-\nu')) + q^n}
{(1-2(-\sqrt{q})^n)^2}
\right], \label{expl}
\ee
where $q = \exp [-\frac{\pi^2}{\tau}]$ and $\nu-\nu' = -\frac{1}{2}\ln c$ . 
The form (\ref{expl}) of the Green function is periodic in the insertion 
coordinates 
$\nu$'s  with a period double (4 instead of 2) with respect to the annulus
case: this is  because the boundary of the M\"{o}bius strip
can be viewed as having double length with respect to one of the two
boundaries in the annulus. For our
purposes we will need another form of $G_{\M}(\nu)$, first proposed by 
Fradkin and Tseytlin \cite{FT}.  Using \be
\ln[1+b^2-2\,b\,\cos x] = -2\sum_{n=1}^{\infty} \frac{b^n}{n} \,\cos \,nx
\nonumber
\ee
and
\be
\sum_{n=1}^{\infty}b^n = \frac{b}{1-b}, \nonumber
\ee
 we obtain
\be 
G_{\M}(\nu-\nu') = - \sum_{n=1}^{\infty} \frac{1}{n}
\,\cos \left(\frac{\pi n (\nu-\nu')}{\tau}\right)
                \left[ \frac{1+(-\sqrt{q})^n}{1-(-\sqrt{q})^n} \right],
\ee
where we have used the regularization 
$\sum_{n=1}^{\infty}1 = -\frac{1}{2}$ 
and we have neglected the terms that do not depend on $\nu$. The 
effect of this regularization is that no negative powers of $\alpha'$ 
are generated in the integration over the variables $\nu$'s and 
$\tau$, \cite{VMLRM}. In this way we can replace the exponentials of the Green function simply by an infrared cutoff
and extract from the amplitude only the terms proportional to
$(\alpha')^{2-D/2}$. Keeping this fact in mind we rewrite the amplitude 
(\ref{1lampl})
as  
\be A^{(1)}(p_1, \dots , p_M) = 
\frac{1}{2}\chi_{M} f_N^{a_1,a_2, \dots , a_M }    
 \frac{g_D^M}{(4\pi)^{D/2}} (2\alpha')^{2-D/2}      
 \int_{0}^{\infty}\D^{\M}
\tau I^{(1)}_{M}(\tau) \ee 
where 
\be
I^{(1)}_{\M}(\tau) & =  & (2\alpha')^{-2}\int_{0}^{\tau}d\nu_M  
                     \int_{0}^{\nu_M}d\nu_{M-1} \dots  
                     \int_{0}^{\nu_3}d\nu_2  \nonumber \\ 
     &\times &    \exp\left[\sum_{r < s}p_r G_\M(\nu_{rs})p_s\right]
               \0 \nonumber \\
         &\times &      
       \exp\left[\sum_{r\neq s}\left(p_r\d_s G_\M(\nu_{sr})\epsilon_s
      +\frac{1}{2} \epsilon_r\d_r\d_s G_\M(\nu_{sr})\epsilon_s\right) 
\right]_{\rm m.l.} 
\ee
and 
\be
\D^{\M} \tau = d\tau w^{-1} \tau^{-D/2} 
\prod_{n=1}^{\infty}(1-(-w)^n)^{2-D}
\ee
Going to the variables $\hat{\nu} = \nu / \tau$ it is easier to 
implement the non--orientability of the M\"{o}bius band. We noticed 
above that the Green function  $G^{\M}$ has double period in $\hat{\nu}$. The
integration region must be chosen accordingly: the 
integration range is now $[0,2]$ instead of $[0,1]$, because we need 
to make two complete revolutions to go around the boundary back to the 
starting point.
\be 
I^{(1)}_{\M}(\tau) & = & (2\alpha')^{-2}\tau^{M-1}\int_{0}^{2}d
\hat{\nu}_M \int_{0}^{\hat{\nu}_M}        d \hat{\nu}_{M-1} \dots
\int_{0}^{\hat{\nu}_3}d\hat{\nu}_2  
\nonumber \\
&\times &   
 \exp \left[\sum_{r
< s}p_r G_\M(\hat{\nu_{rs}})p_s\right] \nonumber \\      
&\times & \exp \left[\sum_{r\neq s}\left(p_r
\frac{1}{\tau}\hat{\d}_s G_\M(\hat{\nu_{sr}})\epsilon_s  +\frac{1}{2}
\frac{1}{\tau^2} \epsilon_r\hat{\d}_r\hat{\d}_s G_\M(\hat{\nu_{sr}})\epsilon_s
\right)
\right] 
\ee 
For the two point function, after a partial integration with
null boundary terms, we obtain 
\be I^{(1)}_2 & = & \epsilon_1 \cdot \epsilon_2
p_1 \cdot p_2 \tau \int_{0}^{2} d\hat{\nu} 
\left(\frac{1}{\tau}{\hat{\d}}G_\M
(\hat{\nu})\right)^2 e^{2\alpha'p_1\cdot p_2 G_\M (\hat{\nu})} \\           
& = & \epsilon_1 \cdot \epsilon_2 p_1 \cdot p_2 \int_{0}^{2}
\left[\sum_{m=1}^{\infty}\frac{\pi }{\tau}\sin (\pi m \hat{\nu})   
\left[\frac{1+(-\sqrt{q})^m}{1-(-\sqrt{q})^m} \right]       \right]
\nonumber\\           &
\times & \left[\sum_{n=1}^{\infty}\frac{\pi}{\tau}\sin (\pi n \hat{\nu})
\left[ \frac{1+(-\sqrt{q})^n}{1-(-\sqrt{q})^n} \right]       \right] 
\ee 
The partial integration has yielded the appropriate powers of 
$\alpha'$, so we can disregard the exponentials of the Green functions,
and perform the $\hat{\nu}$ integration with the help of the formula  
\be
\int_{0}^{2}dx \,\sin(\pi nx)\, \sin(\pi mx) = \delta_{nm} \label{formula}
\ee 
and we
are left with 
\be I^{(1)}_2 = \epsilon_1 \cdot \epsilon_2 \,p_1 \cdot p_2
\frac{\pi^2}{\tau}
\sum_{m=1}^{\infty}\left(\frac{1+(-\sqrt{q})^m}{1-(-\sqrt{q})^m}  
\right)^2
\ee 
Since the integration over $\tau$ will be shared by the 3-- and 4--point
functions, let us define \be Z_{\M} = \pi^2
\int^{\infty}_{0}\frac{\D^{\M}\tau}{\tau}\sum_{m=1}^{\infty}
\left(\frac{1+(-\sqrt{q})^m}{1-(-\sqrt{q})^m}   \right)^2 
\ee 
The sum present in $Z_{\M}$ can be rewritten as  
\be
\sum_{n=1}^{\infty} 
\left(\frac{1+(-\sqrt{q})^m}{1-(-\sqrt{q})^m}   \right)^2 = 
-4(-\sqrt{q}) \frac{d}{d(-\sqrt{q})}\ln \left[(-\sqrt{q})^{1/8}\prod_{n=1}^{\infty}(1-(-\sqrt{q})^n)]\right]~,
\ee
then, using the relation (8.A.27) of \cite{GSW}, we can go to the $k$ 
representation which is more suitable for the field theory limit  
\be
f(-\sqrt{q}) & = & \prod_{n=1}^{\infty} (1-(-\sqrt{q})^n) \nonumber \\ 
 & = & w^{1/24}q^{-1/48} 
\left(- \frac{\ln w}{\pi}\right)^{1/2} f(-w) \nonumber \\  
 & = &  w^{1/24}q^{-1/48} 
\left(- \frac{\ln w}{\pi}\right)^{1/2} \prod_{n=1}^{\infty} (1-(-w)^n).
\ee
and find the following expression for $Z_{\M}$
\be
Z_{\M} & = & \pi^2 \int_{0}^{\infty} \frac{\D^{\M}\tau}{\tau} 
4\sqrt{q}\frac{d}{d(-\sqrt{q})} \ln \left[(-\sqrt{q})^{1/8}\left(-\frac{\ln
w}{\pi}\right)w^{1/24}q^{-1/48} f(-w)    \right] \nonumber \\   & =  & 4
\int_{0}^{\infty}\frac{\D^{\M}\tau}{\tau} w (\ln w)^2  \left[
-\frac{\pi^2}{12}\frac{1}{w(\ln k)^2} + \frac{1}{2w}\frac{1}{\ln w} +
\frac{1}{24} + \sum_{n=1}^{\infty}\frac{n(-w)^{n-1}}{(1-(-w)^n)}\right] 
\nonumber  
\ee
Now we expand the partition function present in $\D^{\M}\tau$ in powers of $w =
e^{-2\tau}$, and keep only the power  $\tau^{1-D/2}$, that is the only one
that gives rise to divergences in the dimensional regularization  
\be 
Z_{\M} =
\frac{2}{3} (26-D) \int_{0}^{\infty} d\tau \tau^{1-D/2} 
e^{-2\alpha'm^2\tau} =
\frac{2}{3} (26-D) \Gamma \left(2-\frac{D}{2}\right) (2\alpha' m^2)^{D/2-2}
\nonumber
\ee 
After setting $\epsilon = 2-D/2$, we obtain for the two point function
\be 
\left. A^{(1)}_{\M}(p_{1},p_{2})\right|_{\rm div} &  = & \frac{i}{2}\tr(t^{a_1}t^{a_2})
\frac{g_D^2}{(4\pi)^{D/2}}  (2\alpha')^{2-D/2}
(2\alpha' m^2)^{D/2-2} \nonumber \\ 
 & \times & \epsilon_1 \cdot \epsilon_2 \, p_1 \cdot p_2
\frac{2}{3} (26-D) \Gamma \left(2-\frac{D}{2}\right) \nonumber \\ 
 & = & \frac{g^2}{(4\pi)^2}\,\frac{11}{3}\,\frac{1}{\epsilon}A^{(0)}(p_1,p_2)
\ee
where $g\equiv g_4$.
For three gluons we have 
\be
I^{\M}_{3} = \frac{1}{\tau} \int_{0}^{2} d \hat{\nu}_3 
\int_{0}^{\hat{\nu}_{3}} d \hat{\nu}_{2}
\left\{ -\epsilon_1 \cdot \epsilon_2 \hat{\d}_2^2 G(\hat{\nu}_2)
\left[ p_1\cdot\epsilon_3 \hat{\d}_3 G(\hat{\nu}_3) + p_2 \cdot \epsilon_3 
\hat{\d}_3 G(\hat{\nu}_{32})   \right]+ \dots         \right\} \0
\ee
where the dots stand for the terms obtained by cyclic symmetry and 
for terms of higher order in $\alpha'$. The  power of $\alpha'$ in the
expression above  is the correct one, without partial integration: also in
this case we can neglect the exponentials, because they are irrelevant 
for ultraviolet divergencies. The integral over $\hat{\nu}$'s coordinates is done again using the formula (\ref{formula}):  
\be
I^{\M}_{3} = 2 \Big( \epsilon_1 \cdot \epsilon_2 
p_2\cdot \epsilon_3 +  \epsilon_2 \cdot \epsilon_3  p_3\cdot \epsilon_1 + 
 \epsilon_1 \cdot \epsilon_3  p_1\cdot \epsilon_2 \Big)
\frac{\pi^2}{\tau}
\sum_{n=1}^{\infty} 
\left(\frac{1-(-\sqrt{q})^{n}}{(1-(-\sqrt{q})^n)}\right)^2 \nonumber
\ee
\be
\left. A^{(1)}_{\M}(p_{1},p_{2},p_3)\right|_{\rm div}
 & = & \frac{1}{2}\tr(t^{a_1}t^{a_2}t^{a_3}) 
      \frac{g_D^2}{(4\pi)^{D/2}} (2\alpha')^{2-D/2}(2\alpha' m^2)^{D/2-2} 
\nonumber\\
 & \times &
2\Big( \epsilon_1 \cdot \epsilon_2 
p_2\cdot \epsilon_3 +  \epsilon_2 \cdot \epsilon_3  p_3\cdot \epsilon_1 + 
 \epsilon_1 \cdot \epsilon_3  p_1\cdot \epsilon_2 \Big) 
 \frac{2}{3} (26-D) \Gamma \left(2-\frac{D}{2}\right) \0 \\ 
 & = & 
\frac{g^2}{(4\pi)^2}\,\frac{11}{3}\,\frac{1}{\epsilon}A^{(0)}(p_1,p_2,p_3)
\ee
Finally for four gluons we have
\be
I^{\M}_{4} & = & \frac{1}{\tau^2}\int_{0}^{2} d \hat{\nu}_4 
\int_{0}^{\hat{\nu}_{4}} d \hat{\nu}_3
\int_{0}^{\hat{\nu}_3} d\hat{\nu}_2
\Big( \epsilon_1 \cdot \epsilon_2 \epsilon_3 \cdot \epsilon_4 
\hat{\d}^2_2 G(\hat{\nu}_2) \hat{\d}^2_4 G(\hat{\nu}_{43})  \nonumber \\
 & + &
\,\epsilon_1 \cdot \epsilon_3 \epsilon_2 \cdot \epsilon_4 
\hat{\d}^2_3 G(\hat{\nu}_3) \hat{\d}^2_4 G(\hat{\nu}_{42})  \nonumber \\
 & + &
\,\epsilon_1 \cdot \epsilon_4 \epsilon_3 \cdot \epsilon_2 
\hat{\d}^2_4 G(\hat{\nu}_4) \hat{\d}^2_3 G(\hat{\nu}_{32}) + \dots 
\Big) 
\ee
Again we have the correct power of $\alpha'$ without partial 
integration and we can discard the exponential; the dots denotes terms
proportional to the external momenta that will play no role because they 
are not present in the 1PI tree level diagrams.  
\be 
I^{\M}_{4} & = & 
2\left(-\frac{1}{2}  
\epsilon_1 \cdot \epsilon_2 \epsilon_3 \cdot \epsilon_4  + \epsilon_1 \cdot
\epsilon_3 \epsilon_2 \cdot \epsilon_4  -\frac{1}{2} \epsilon_1 \cdot
\epsilon_4 \epsilon_3 \cdot \epsilon_2 \right)  
\frac{\pi^2}{\tau} \sum_{n=1}^{\infty}
\left(\frac{1+(-\sqrt{q})^{n}}{1-(-\sqrt{q})^n}\right)^2 \0
\ee
Therefore
\be
\left. A^{(1)}_{\M}(p_{1},p_{2},p_3,p_4)\right|_{\rm div}
 & = & \frac{i}{2}\tr(t^{a_1}t^{a_2}t^{a_3}t^{a_4}) 
      \frac{g_D^2}{(4\pi)^{D/2}} (2\alpha')^{2-D/2}(2\alpha' m^2)^{D/2-2} 
 \0 \\
 & \times & 2\left(-\frac{1}{2}   \epsilon_1 \cdot \epsilon_2 \epsilon_3 
\cdot 
\epsilon_4 
+ \epsilon_1 \cdot \epsilon_3 \epsilon_2 \cdot \epsilon_4 
-\frac{1}{2} \epsilon_1 \cdot \epsilon_4 \epsilon_3 \cdot \epsilon_2 \right)
 \frac{2}{3} (26-D) \Gamma \left(2-\frac{D}{2}\right) \0 \\
 & = & \frac{g^2}{(4\pi)^2}\,\frac{11}{3}\,\frac{1}{\epsilon}
A^{(0)}(p_1,p_2,p_3,p_4)
\ee 
We can now summarize our results by collecting together the planar 
amplitudes (\ref{A1A0}) and the M\"{o}bius ones. The final result is

\be
 \left. A^{(1)}(p_1,...)\right|_{\rm div}=
-\frac{N-2}{2}\, \frac{g^2}{(4\pi)^2} \,\frac {11}{3} 
 \frac
{1}{\epsilon}A^{(0)}(p_1,...)\label{A1A0M}
\ee

\subsection{M\"{o}bius amplitudes: second method}

The second method is more laborious, but it
has the advantage that one can single out more explicitly the regions of the moduli 
space corresponding to the different divergent contributions and thus provides 
a better understanding of the field theory limit. The
$\a'\to 0$ limit corresponds to the parameters $\tau$ and $\nu_r$ going to infinity,
or, more precisely, to $\tau\to\infty$ and $\hat \nu_r= \frac {\nu_r}{\tau}$ finite. 
Therefore, on the basis of (\ref{1lampl}), we need the corresponding asymptotic 
expansion of $G_\M(\nu)$ and its derivatives. The latter is given, up to
${\cal O}(e^{-4\tau})$ terms, by (from now on we drop the subscript $_\M$
from the propagator):
\be
G^+(\nu) &=& - \hat \nu^2 \tau + \hat\nu\tau -e^{-2 \hat\nu\tau }+ e^{-2\tau}
\left(e^{-2 \hat\nu\tau }+e^{2 \hat\nu\tau }-1\right)\0\\
\d_\nu G^+(\nu) &=& - 2\hat \nu+ 1 +2e^{-2 \hat\nu\tau }+ 2e^{-2\tau}
\left(e^{2 \hat\nu\tau }-e^{-2 \hat\nu\tau }\right)\label{asG+}\\
\d^2_\nu G^+(\nu) &=& - \frac {2}{\tau}-4e^{-2 \hat\nu\tau }+ 4e^{-2\tau}
\left(e^{2 \hat\nu\tau }+e^{-2 \hat\nu\tau }\right)\0
\ee
and
\be
G^-(\nu) &=& - \hat \nu^2 \tau + \hat\nu\tau +e^{-2 \hat\nu\tau }- e^{-2\tau}
\left(e^{-2 \hat\nu\tau }+e^{2 \hat\nu\tau }+1\right)\0\\
\d_\nu G^-(\nu) &=& - 2\hat \nu+ 1 -2e^{-2 \hat\nu\tau }-2e^{-2\tau}
\left(e^{2 \hat\nu\tau }-e^{-2 \hat\nu\tau }\right)\label{asG-}\\
\d^2_\nu G^-(\nu) &=& - \frac {2}{\tau}-4e^{-2 \hat\nu\tau }- 4e^{-2\tau}
\left(e^{2 \hat\nu\tau }+e^{-2 \hat\nu\tau }\right)\0
\ee

To compute the one--loop amplitude we have to specify which partial 
propagator $G^+$ or $G^-$ we have to insert in eq.(\ref{1lampl}). To this 
end we split the boundary of $\M$ into two parts $AA'$ lying in the
positive real $\rho$ axis, and $BB'$ along the negative $\rho$ axis 
(see figure). One has to consider all the configurations which are compatible
with any given ordering of the gluon insertions along the boundary of
$\M$.

\subsubsection{Two--gluon amplitude}

In the two--gluon amplitude only one propagator is involved. Therefore
the two--gluon amplitude on $\M$ contains two contributions, one with 
$G^+$ corresponding to the gluon insertions in the same interval $AA'$ or 
$BB'$ and the other with $G^-$ corresponding to one insertion in
$AA'$ and the other in $BB'$. We will use translational invariance in order
to fix the insertion 1 at the point $A'$, i.e. $\rho_1=1$ or $\nu_1=0$.
After these preliminaries we insert all the data in eq.(\ref{1lampl}) and 
find
\be
A_\M^{(1)}(p_1,p_2) &=& \frac i2\tr (t^{a_1}t^{a_2})  
\frac {g_D^2}{(4\pi)^{\frac D2}}(2\alpha')^{2-\frac D2}
\,\int_0^\infty d\tau e^{2\tau}\tau^{-\frac D2}\prod_{n=1}^\infty 
\left(1-(-1)^ne^{-2n\tau}\right)^{2-D}\0\\
& \times & (- \epsilon_1\cdot \epsilon_2)\int_0^\tau d\nu \left(
 e^{2\alpha' p_1\cdot p_2 G^+(\nu)} \d^2_\nu G^+(\nu) +
e^{2\alpha' p_1\cdot p_2 G^-(\nu)}  
\d^2_\nu G^-(\nu)\right)
 \label{1l2p} 
\ee 
where $\nu=\nu_2$.

Now we integrate by parts in $\nu$ and disregard the contributions at 
$\nu = 0,\tau$, since, as was noticed in \cite{VMLRM}, they correspond
in field theory to massless tadpole contributions, which are defined to 
vanish in dimensional regularization. Therefore the RHS of (\ref{1l2p})
can be replaced by:
\be 
&&\frac i2\tr (t^{a_1}t^{a_2})  
\frac {g_D^2}{(4\pi)^{\frac D2}}(2\alpha')^{2-\frac D2}
\,\int_0^\infty d\tau e^{2\tau}\tau^{-\frac D2}\prod_{n=1}^\infty 
\left(1-e^{-2n\tau}\right)^{2-D}\0\\
&&  (\epsilon_1\cdot \epsilon_2\,p_1\cdot p_2)\int_0^\tau d\nu \left(
 e^{2\alpha' p_1\cdot p_2 G^+(\nu)} \left(\d_\nu G^+(\nu)\right)^2 +
e^{2\alpha' p_1\cdot p_2 G^-(\nu)}  
\left(\d_\nu G^-(\nu)\right)^2\right)\label{1l2p'}
\ee

At this point we insert the expansions (\ref{asG+}) and (\ref{asG-}) and 
evaluate the $\nu$ integral first. One notices that the two exponentials
$e^{2\alpha' p_1\cdot p_2 G^\pm(\nu)}$ for large $\tau$ can be written
as $e^{2\alpha' p_1\cdot p_2 \,(\hat \nu-\hat \nu^2)\tau}$ and play
the role of a cutoff factor. Therefore, for large $\tau$, the $\nu$ integral
in (\ref{1l2p'}) is determined by
\be
\int_0^\tau d\nu \,2\,\left( (1-2 \hat \nu)^2+8 e^{-2\tau}\right)
e^{2\alpha' p_1\cdot p_2 \,(\hat \nu-\hat \nu^2)\tau}\0
\ee
Now inserting this equation back into (\ref{1l2p'}), we see that there are
contributions to the $\tau$ integral proportional to $e^{2\tau}$. These
are recognized to be contributions from the tachyon and must be discarded 
by hand (this ad hoc operation is the price we have to pay for having
embedded our gauge theory in the bosonic string rather than in a superstring
theory). The terms of zeroth order in $e^{2\tau}$ are the relevant ones
for our purposes. As shown in \cite{VMLRM}, these integrals can be exactly
evaluated and the pole in $\epsilon= \frac {4-D}{2}$ easily extracted.
The result is
\be
\left. A_\M^{(1)}(p_1,p_2)\right|_{\rm div} &=&i\tr (t^{a_1}t^{a_2})  
\frac {g^2}{(4\pi)^2}\epsilon_1\cdot \epsilon_2\,p_1\cdot p_2
\frac {11}{3} \frac {1}{\epsilon}\label{2pdiv}
\ee
which, if we forget the factor of $N$, is twice the planar contribution
with opposite sign. If we put together the 
results for the planar annulus amplitude and the M\"{o}bius strip we finally
obtain for the 1PI divergent part of the two--gluon amplitude
\be
\left. A^{(1)}(p_1,p_2)\right|_{\rm div} &=&- \frac {N-2}{2}   
\frac {g^2}{(4\pi)^2}\frac {11}{3}\frac {1}{\epsilon}
A^{(0)}(p_1,p_2)\label{2ptotdiv}
\ee
This is exactly what is expected from renormalization 
theory in the background field method formalism.

\subsubsection{Three--gluon amplitude}  

The three--gluon amplitude involves two propagators and four possible
configurations for any given ordering of the external legs, \cite{GNSS}. 
The
four configurations can be  classified as follows. The orientation of the
boundary of $\M$ is chosen from $A'$ to $A$ and from $B$ to $B'$. We call it
the standard orientation. We  consider
the three insertions at $\rho_1,\rho_2,\rho_3$ ordered according to
the standard orientation and set $\rho_1=A'$, see figure. Now we append by
convention a + or a -- to $\r$ according to whether $\r$ falls in the
interval $AA'$ or in $BB'$. The four configurations are then specified
as follows  
\begin{itemize}
\item s1: ($\r_1^+, \r_2^+, \r_3^+$)

\item s2: ($\r_1^+, \r_2^+, \r_3^-$)

\item s3: ($\r_1^+, \r_2^-, \r_3^-$)

\item s4: ($\r_1^+, \r_3^-, \r_2^+$)

\end{itemize}

Each triple is in order of decreasing modulus. For instance, s4 means
$|\r_1|\geq|\r_3|\geq|\r_2|$ and that $\r_1$ and $\r_2$ are in $AA'$
while $\r_3$ is in $BB'$. s1--s4 specify distinct sectors of the integration
region (moduli space).

The amplitude given by (\ref{1lampl}) contains three pieces, which
are proportional to the three terms contained in the RHS of (\ref{3p0}).
We will consider here the one proportional to $\epsilon_1\cdot\epsilon_2
\,p_2\cdot \epsilon_3$. The corresponding coefficient in 
$A^{(1)}(p_1,p_2,p_3)$ is given by
\be
&& \frac 12 \tr (t^{a_1}t^{a_2}t^{a_3})
\frac {g_D^3}{(4\pi)^{\frac D2}}(2\alpha')^\frac{4-D}{2}
\,\int_0^\infty d\tau e^{2\tau}\tau^{-\frac D2}\prod_{n=1}^\infty 
\left(1-(-1)^ne^{-2n\tau}\right)^{2-D}\,  \label{c1c2c3c4}\\
&&\left\{\int_0^{\tau}d\nu_3 \int_0^{\nu_3}d\nu_2 \, 
e^{2\a'\left[p_1\cdot p_2 G_+(\nu_2) +p_2\cdot p_3 G_+(\nu_{32})+
p_1\cdot p_3 G_+(\nu_3)\right]} \d_{\nu_2}^2G_+(\nu_2)
\d_{\nu_3}\left(G_+(\nu_{32})- G_+(\nu_{3})\right)\right.\0\\
&& +\int_0^{\tau}d\nu_3 \int_0^{\nu_3}d\nu_2 \, 
e^{2\a'\left[p_1\cdot p_2 G_+(\nu_2) +p_2\cdot p_3 G_-(\nu_{32})+
p_1\cdot p_3 G_-(\nu_3)\right]} \d_{\nu_2}^2G_+(\nu_2)
\d_{\nu_3}\left(G_-(\nu_{32})- G_-(\nu_{3})\right)\0\\
&& +\int_0^{\tau}d\nu_3 \int_0^{\nu_3}d\nu_2 \, 
e^{2\a'\left[p_1\cdot p_2 G_-(\nu_2) +p_2\cdot p_3 G_+(\nu_{32})+
p_1\cdot p_3 G_-(\nu_3)\right]} \d_{\nu_2}^2G_-(\nu_2)
\d_{\nu_3}\left(G_+(\nu_{32})- G_-(\nu_{3})\right)\0\\
&& \left. +\int_0^{\tau}d\nu_2 \int_0^{\nu_2}d\nu_3 \, 
e^{2\a'\left[p_1\cdot p_2 G_+(\nu_2) +p_2\cdot p_3 G_-(\nu_{32}
)+
p_1\cdot p_3 G_-(\nu_3)\right]} \d_{\nu_2}^2G_+(\nu_2)
\d_{\nu_3}\left(G_-(\nu_{32})- G_-(\nu_{3})\right)\right\} \0
\ee
The last four lines in this equation correspond to the contributions 
from the four configurations listed above, in the same order.
As analyzed in \cite{VMLRM}, the divergent contributions corresponding to 
1PI diagrams in the $\a'\to 0$  limit, come from two different regions of 
the moduli space, which we call type I and type II. 

The type I region corresponds to the three 
insertion points being kept widely separated while $\tau\to\infty$, i.e.
while the M\"{o}bius strip shrinks to zero size ($w\to 0, q\to 1$). Intuitively,
this corresponds in the field theory language to Feynman diagrams
with three propagators and three three--point vertices.
This means that $\nu_3$ and $\nu_{32}$ are of order $\tau$ while
$\tau\to\infty$. It is possible to show that these contributions
come only from the first terms (those not containing exponentials)
in the asymptotic expansions (\ref{asG+}, \ref{asG-}).
We seem to have four contributions of this type, corresponding to the four
configurations s1--s4. However this is not the case. Only two of them
contribute to type I, precisely s1 and s3. In s2 and s4, point 2 and 
point 3 are bound to lie on opposite sides of the band; in the field theory 
limit these contributions do not flow toward the expected Feynman diagrams.
In a sense they are analogous to the nonplanar ones. 
 
To evaluate the type I contribution we remark that the exponentials in 
eq.(\ref{c1c2c3c4}) play simply the role of dumping factors. Therefore 
we simplify things by replacing them with a universal dumping factor 
$e^{-2\a' m^2\tau}$. After discarding the tachyon contribution one can see
that the relevant UV divergent part from region of type I in
eq.(\ref{c1c2c3c4}) is contained in  
\be
\frac 12\tr (t^{a_1}t^{a_2}t^{a_3})
\frac {g_D^3}{(4\pi)^{\frac D2}}(2\alpha')^\frac{4-D}{2}
\,\int_0^\infty d\tau \tau^{-\frac D2}(2-D)
\int_0^{\tau}d\nu_3 \int_0^{\nu_3}d\nu_2\, e^{-2\a' m^2\tau}
 8 \left( \frac{\nu_2}{\tau^2}\right)\0
\ee
After a standard integration, this
 becomes
\be
- \tr (t^{a_1}t^{a_2}t^{a_3})\frac {g_D^3}{(4\pi)^\frac D2} 
\frac 43
m^{D-4}
 \Gamma(\epsilon)\0
\ee
Collecting the above results and setting $D=4$ one finds the type I
contribution to the divergent part of the three--gluon amplitude is:
\be
\left. A_\M^{(1)}(p_1,p_2,p_3)\right\vert_{I}=-\tr (t^{a_1}t^{a_2}t^{a_3})\,
\frac {g^3}{(4\pi)^2}\,
\epsilon_1\cdot\epsilon_2\,p_2\cdot \epsilon_3 \, \frac 43 \,
\frac {1}{\epsilon}\label{3pI}
\ee
 
Let us pass now to the type II region. It is the region in the moduli
space where two insertion points come close together like $1/\tau$ as
$\tau\to\infty$. In field theory such terms correspond to one--loop
three--gluon diagrams with one four--point vertex. There are three
possibilities: either $\r_1\to 1$, or $\r_1\to\r_2$, or $\r_3\to -w$.
These correspond to either $\hat\nu_2 \sim {\cal O}(\tau^{-1})$ or  
$\hat\nu_{32} \sim {\cal O}(\tau^{-1})$ or 
$\hat\nu_2 \sim {\cal O}(\tau^{-1})$. In field theory terms this
corresponds to Feynman diagram with two internal propagators and one
four--point vertex.

Using the asymptotic expansions (\ref{asG+}, \ref{asG-}) into 
(\ref{c1c2c3c4}) one can see that the type II contributions 
can only come from the exponential terms in (\ref{asG+}, \ref{asG-}).
Once again, however, we should not apply the formulas mechanically.
The type II contributions of the sectors s1--s4 must be
carefully evaluated. For instance it is evident that in s4 the
punctures 2 and 3 cannot approach each other because they are confined to lie
on
 opposite sides of the band. On the other hand $\r_3$ cannot go to 
$-w$ because $|\r_3|\geq |\r_2|$, and, for the same reason $\r_2$ cannot 
go to 1. Therefore neither 3 nor 2 can get close to 1. Thus sector s4
is not going to contribute to type II. On the other hand, in s1
we have the possible collapses $2\to 1$ and $2\to3$,
in s2 we have the only possible collapse $3\to 1$, while in s4 we can have
both $3\to1$ and $2\to 3$. As it turns out, $2\to 1$ does not contributes
to the divergent part. Carrying out the explicit calculations, the divergent
part of (\ref{c1c2c3c4}), as far as type II is concerned, is contained in 
\be
\frac 12\tr (t^{a_1}t^{a_2}t^{a_3})&&
\frac {g_D^3}{(4\pi)^{\frac D2}}(2\alpha')^\frac{4-D}{2}
\,\int_0^\infty d\tau \tau^{-\frac D2} e^{2\tau}
\int_0^{\tau}d\nu_3 \int_0^{\nu_3}d\nu_2\, e^{-2\a' m^2\tau}\0\\
&&\left[\, 8 e^{-2\tau +2(\nu_3-\nu_2)} - 8^{-2\tau +2(\nu_2-\nu_3)}\right.\0\\
&&\! + \, 8 e^{-2\nu_3}-8 e^{-4\tau+2\nu_3}\0\\
&&\left. \! +\, 8 e^{-2\tau +2(\nu_3-\nu_2)} - 8^{-2\tau +2(\nu_2-\nu_3)}
+ 8 e^{-2\nu_3}-8 e^{-4\tau+2\nu_3}\, \right]\label{s1s2s3}
\ee
where the last three lines correspond to the s1, s2 and s3 contributions, 
respectively. The calculation now is straightforward.
Setting $D=4-2\epsilon$ one finds that the type II contribution
to the divergent part of the three--gluon amplitude is:
\be
\left. A_\M^{(1)}(p_1,p_2,p_3)\right\vert_{II}=\tr (t^{a_1}t^{a_2}t^{a_3})\,
\frac {g^3}{(4\pi)^2}\,
\epsilon_1\cdot\epsilon_2\,p_2\cdot \epsilon_3 \, 16 \,
\frac {1}{\epsilon}\label{3pII}
\ee
Finally the total divergent part for the three--gluon amplitude is
\be
\left. A_\M^{(1)}(p_1,p_2,p_3)\right\vert_{\rm I+II}=\tr
(t^{a_1}t^{a_2}t^{a_3})\, \frac {g^3}{(4\pi)^2}\,
\epsilon_1\cdot\epsilon_2\,p_2\cdot \epsilon_3 \, \frac {44}3\,
\frac {1}{\epsilon}\label{3pII'}
\ee
Therefore
\be
\left. A^{(1)}(p_1,p_2,p_3)\right\vert_{\rm I+II}
=- \frac {N-2}{2}   
\frac {g^2}{(4\pi)^2}\frac {11}{3}\frac {1}{\epsilon}
A^{(0)}(p_1,p_2,p_3)\label{3ptotdiv}
\ee

\subsubsection{Four--gluon amplitude}  

The four--gluon amplitude involves three propagators and eight possible
configurations for any given ordering of the external legs, see \cite{GSW}.
The eight configurations can be  classified as above. We consider
the four insertions at $\rho_1,\rho_2,\rho_3,\rho_4$ ordered according to
the standard orientation of the boundary of $\M$ and set $\r_1=A'$.
The corresponding eight sectors of integration are then specified as follows  

\begin{itemize}
\item s1: ($\r_1^+, \r_2^+, \r_3^+,\r_4^+$)

\item s2: ($\r_1^+, \r_2^+, \r_3^+,\r_4^-$)

\item s3: ($\r_1^+, \r_2^+, \r_3^-,\r_4^-$)

\item s4: ($\r_1^+, \r_2^-, \r_3^-,\r_4^-$)

\item s5: ($\r_1^+, \r_4^-, \r_2^+,\r_3^-$)

\item s6: ($\r_1^+, \r_4^-, \r_2^+,\r_3^+$)

\item s7: ($\r_1^+, \r_2^+, \r_4^-,\r_3^-$)

\item s8: ($\r_1^+, \r_3^-, \r_4^-,\r_2^+$)

\end{itemize}
Each quadruple is written in order of decreasing modulus. 

Now we single out in (\ref{1lampl}) the piece proportional to 
$\epsilon_1\cdot\epsilon_3\, \epsilon_2\cdot\epsilon_4$ (the other two pieces
can be dealt with similarly, see \cite{VMLRM}) and simplify the resulting
expression as in the three--gluon case. In particular we replace the
exponential factors with a unique dumping factor $e^{-2 \a' m^2\tau}$.

Next we discus the contributions from region I and II.
To this end we avoid explicitly writing down encumbering equations.  
Let us recall that type I contributions come from well separated
configurations of the punctures in the limit $\tau\to \infty$, they correspond
in field theory to Feynman diagram with four internal propagators. The only two
sectors that can contribute are s1 and s4. All the other sectors are
non--planar--like in that they contain at least two points on opposite sides
of the band. Their field theory limit is different from that expected for 
type I contributions.

As for type II contributions they correspond to two separate couples 
of points coming simultaneously 
together like ${\cal O}(1/\tau)$ as $\tau\to\infty$. In field theory
this correspond to Feynman diagram with two internal propagators. Sector by
sector we find: in s1 we can have $2\to 1$ and $3\to 4$;
in s2 we can have $2\to 3$ and $\r_4\to -w$, i.e. $4\to 1$; in s3 we can 
have $2\to 1$ and $3\to 4$; in s4 we can have $2\to 3$ and $\r_4\to -w$, 
i.e. $4\to 1$; no two separate couples of points can come simultaneously
together in the remaining sectors. So sectors s5--s8 do not contribute 
neither to type  I nor to  type II divergences. 

Now, going to explicit formulas, we find that the relevant multiplicative
factor of $\epsilon_1\cdot\epsilon_3\, \epsilon_2\cdot\epsilon_4$  in
$A^{(1)}(p_1,p_2,p_3,p_4)$ is
\be
&&\frac i2\tr (t^{a_1}t^{a_2}t^{a_3}t^{a_4})
\frac {g_D^4}{(4\pi)^{\frac D2}}(2\alpha')^\frac{4-D}{2}
\, \int_0^\infty d\tau e^{2\tau}\tau^{-\frac D2}\prod_{n=1}^\infty 
\left(1-(-1)^ne^{-2n\tau}\right)^{2-D}\0\\
&&\int_0^{\tau}d\nu_4\int_0^{\nu_4}d\nu_3\int_0^{\nu_3}d\nu_2\, e^{-2\a'
m^2\tau}\Big[\d_{\nu_3}^2 G_+(\nu_3) \d_{\nu_4}^2 G_+(\nu_{42})\0\\
&&\, \,+\d_{\nu_3}^2 G_+(\nu_3) \d_{\nu_4}^2 G_-(\nu_{42})
+\d_{\nu_3}^2 G_-(\nu_3) \d_{\nu_4}^2 G_-(\nu_{42})
+\d_{\nu_3}^2 G_-(\nu_3) \d_{\nu_4}^2 G_+(\nu_{42})\Big]\label{4p1}
\ee
where the terms in square brackets refer to sector s1 down to s4,
respectively.

It remains for us to evaluate the above integral for type I and II.
As pointed out above the type I contributions
come only from the first terms (those not containing exponentials)
in the asymptotic expansions (\ref{asG+}, \ref{asG-}).
\be
\frac i2\tr (t^{a_1}t^{a_2}t^{a_3}t^{a_4})
\frac {g_D^4}{(4\pi)^{\frac D2}}(2\alpha')^\frac{4-D}{2}
\,\int_0^\infty d\tau \tau^{-\frac D2}(2-D)
\int_0^{\tau}d\nu_4\int_0^{\nu_4}d\nu_3\int_0^{\nu_3}d\nu_2\, 
e^{-2\a' m^2\tau}
 \left(\frac{8}{\tau^2}\right) \0
\ee
Proceeding as above this gives rise to the following divergent part of
the four--point amplitude
\be
\left. A_\M^{(1)}(p_1,p_2,p_3,p_4)\right\vert_I=
-i \tr (t^{a_1}t^{a_2}t^{a_3}t^{a_4})\,\frac {g^4}{(4\pi)^2}\,
\epsilon_1\cdot\epsilon_3\, \epsilon_2\cdot\epsilon_4\, \frac 43\,
\frac 1\epsilon\label{4pI}
\ee
Type II contributions come from the terms containing exponentials in
(\ref{asG+}, \ref{asG-}). From (\ref{4p1}) one gets 
\be
&&-\frac i2\tr (t^{a_1}t^{a_2}t^{a_3}t^{a_4}))
\frac {g_D^3}{(4\pi)^{\frac D2}}(2\alpha')^\frac{4-D}{2}
\,\int_0^\infty d\tau \tau^{-\frac D2} e^{2\tau}
\int_0^{\tau}d_{\nu_4}\int_0^{\nu_4}d_{\nu_3} \int_0^{\nu_3}d\nu_2\, 
e^{-2\a'm^2\tau}\0\\
&&  \left[ 32 e^{-2\tau - 2(\nu_2+\nu_3-\nu_4)} + 
32 e^{-2\tau + 2(\nu_2+\nu_3-\nu_4)} + 
+ 32 e^{2(\nu_2-\nu_3-\nu_4)}+ 32 e^{-4 \tau -2(\nu_2-\nu_3-\nu_4)}\right]
\0
\ee 
whose evaluation leads to the divergent part
\be
\left. A_\M^{(1)}(p_1,p_2,p_3,p_4)\right\vert_{II}=
i\tr (t^{a_1}t^{a_2}t^{a_3}t^{a_4})\,\frac {g^4}{(4\pi)^2}\,
\epsilon_1\cdot\epsilon_3\, \epsilon_2\cdot\epsilon_4\, 16\,
\frac 1\epsilon\label{4pII}
\ee
Summing type I and type II we get
\be
\left. A_\M^{(1)}(p_1,p_2,p_3,p_4)\right\vert_{I+II}=
i\tr (t^{a_1}t^{a_2}t^{a_3}t^{a_4})\,\frac {g^4}{(4\pi)^2}\,
\epsilon_1\cdot\epsilon_3\, \epsilon_2\cdot\epsilon_4\,\,  \frac {44}3\,
\frac 1\epsilon\label{4pI+II}
\ee
Once again we obtain
\be
\left. A^{(1)}(p_1,p_2,p_3,p_4)\right\vert_{\rm I+II}
=- \frac {N-2}{2}\,   
\frac {g^2}{(4\pi)^2}\,\frac {11}{3}\,\frac {1}{\epsilon}
A^{(0)}(p_1,p_2,p_3,p_4)\label{4ptotdiv}
\ee
Eqs.(\ref{2ptotdiv}, \ref{3ptotdiv}, \ref{4ptotdiv}) coincide with the results
of the previous subsection. We remark that they
are the one--loop 
 quantum corrections expected in an $SO(N)$ gauge field
theory in the
 background field formalism, \cite{abbot,VMLRM}. 
They correspond to a
 renormalization constant 
\be
Z_A= 1  + \frac {N-2}{2} 
\frac {g^2}{(4\pi)^2}\,\frac {11}{3}\,\frac {1}{\epsilon}\label{renconst}
\ee
This amounts to one--loop renormalizability (in 4D) of the low energy 
effective 
action of the string theory with $so(N)$ Chan--Paton factors, that is the
well--known fact that $SO(N)$ gauge field theory in 4D is renormalizable.

\section{The string propagator on a non-orientable world-sheet. 
Case $B\neq 0$}

We turn now to the same problems considered in section 2 and 3, but in 
the presence of a constant $B$ field. A few words of caution are in order.

A D-brane with an $SO(N)$ (or $Sp(N)$) gauge theory on it can be found 
in correspondence with an orientifold: it corresponds to a set of branes
and mirror branes which collapse on the orientifold. This fact entails 
a problem when we want to consider such a system in the presence of a
$B$ field. In fact the orientifold projection contains a space inversion
which seems to exclude the presence of a $B$ field in the final 
configuration. It was however argued in \cite{BSST} that this is not
a cogent difficulty, a way out can be found. Here we add an alternative 
simple argument to the one presented in \cite{BSST}, which seems to be 
more appropriate to the type of problems we consider in this paper. 
In the original (before projection) theory one can always add to the $B$ field a
constant part without changing the equations of motion of (super)gravity.
This constant part is not directly affected by the string oscillators 
(which determine the equations of motion of the low energy effective 
action via the string amplitudes). On the other hand the orientifold projection
operator is defined through the action on the string oscillators, so 
that a constant `relic' $B$ field may conceivably not be affected by the 
projection\footnote{In $SO(N)$there is no global $U(1)$ factor as in
$U(N)$. Therefore one may wonder whether the $B$ field, which is
not protected by the gauge invariant combination $B-dA$, might be gauged 
away. The answer is no, because the $B$ field after the orientifold 
projection is not dynamical anymore, it does not appear in the effective 
action, so also its gauge properties disappear. Said
differently, away from the orientifold every brane has a $U(1)$ field on
it which guarantees the existence of a nonvanishing gauge invariant
combination $B-dA$; it is natural to assume that in 
the collapsing limit, by continuity, the value of the gauge invariant 
combination $B-dA$ will be unchanged even though a (global) $U(1)$ 
$A$ has disappeared.}. For similar considerations, see \cite{BS}. 

In this paper we give all this for granted and consider a set of 
D-branes collapsed over an orientifold with orthogonal (or symplectic) 
Chan--Paton factors in the presence of a constant $B$ field. This is 
expected to give rise to a \nc $SO(N)$ $(Sp(N))$ gauge field 
theory. The tree level analysis of such theories has been carried 
out in \cite{BSST}. As explained above, in this paper we wish 
to do the one--loop analysis. But this entails a new
problem. In fact the sigma--model action for open strings attached to a 
D--brane is (we adopt the conventions of \cite{SW})
 \be
\frac{1}{4\pi \alpha'}\int_\Sigma d^2x \left(\sqrt{h} 
h^{\alpha\beta}\d_\alpha X^i\d_\beta X^j g_{ij}
-2 \pi \alpha' \int_{\Sigma}d^2x \epsilon^{\alpha\beta}\d_\alpha X^i\d_\beta X^j B_{ij}
\right)\label{action}
\ee
where $\Sigma$ is the string world--sheet, $g_{ij}$ is the closed string
metric and $B_{ij}$ are the components of the constant $B$ field. 
At tree level 
the relevant world--sheet is the disk, while at one--loop the relevant
world--sheets are the annulus and the M\"{o}bius band. Disk and annulus 
are orientable and the integrals in (\ref{action}) are well--defined 
on such surfaces. But the M\"{o}bius strip in nonorientable and, 
while the first term in (\ref{action}) is well defined on it, 
the second is not. The reason 
is that, as we have recalled in section 2, on nonorientable manifolds 
only {\it densities} can be integrated,
see \cite{botu}. Now, the first integrand in (\ref{action}) is a 
density, while the second is not (it is the component 
of the pull--back of a two--form). Therefore
the second part of (\ref{action})
is meaningless when $\Sigma$ is the M\"{o}bius band. However, since B is
constant, in general we can replace (\ref{action}) with 
\be
\frac{1}{4\pi \alpha'}\int_\Sigma d^2x \sqrt{h} 
h^{\alpha\beta}\d_\alpha X^i\d_\beta X^j g_{ij}
- \frac i2\int_{\d\Sigma}dt X^i\d_t X^j B_{ij}
\label{actionb}
\ee
where $\d_t$ is the derivative tangent to the boundary $\d\Sigma$.
This expression is now well--defined also for the M\"{o}bius strip since
its boundary (a circle) is orientable. From now on we will use
(\ref{actionb}) instead of (\ref{action}).

Let us turn now to the string propagator.
The problem is to find, on the surface $\Sigma$ of interest
(in our case $\M$), the solution of the equation
\be
4\d_\rho\d_{\bar \rho} \G^{ij}(\rho,\rho')= 
2{\pi} \alpha' g^{ij} \delta(\rho-\rho')\label{greenrhoB}
\ee
and
\be
[(g+F)^i{}_k \rho\d_\rho - (g-F)^i{}_k \bar \rho \d_{\bar\rho}]
\left. \G^{kj}(\rho,\rho')\right\vert_{\d\Sigma}= Kg^{ij}\label{bcrhoB}
\ee
where $K$ is the same as in section 2.
Moreover we require that $\G^{ij}(\r,\r')=\G^{ji}(\r',\r)$.

The solution we propose for the M\"{o}bius strip is as follows.
\be
\frac {1}{\alpha'}\G_\M^{ij}(\rho,\rho') = 
g^{ij}\left(\I_\M(\rho,\rho')+ f_\M(\r,\r')\right) + 
(2{\hat g}^{ij}-g^{ij})\J_\M(\rho,\rho')  +
\frac{\theta^{ij}}{\a'}\K_\M(\rho,\rho')\label{GMB}
\ee
where
\be
\hat g^{ij}= \left(\frac {1} {g+F} g \frac {1}{g-F}\right)^{ij},
\quad\quad \theta^{ij}= -2\pi \a'\left(\frac {1}{g+F }F
\frac{1}{g-F}\right)^{ij}
 \label{gtheta}
\ee
are the open string metric and the deformation parameter, respectively, 
$\I_M, f_\M$ and $\J_M$ are the same as in section 2, and
\be
\K_\M(\rho,\rho') &=& \frac{(\ln\frac{\r} {\rb'})^2-
(\ln\frac{\rb}{\r'})^2}{2\,\ln w}+
\ln\, \frac{\r-\rb'}{\rb-\r'}+\frac 12 \,\ln \frac {\rb\r'}{\r\rb'}\0 \\
&&\0\\
& + & \ln 
\prod_{n=1}^{\infty}\frac{(1-(-w)^n\frac{\r}{\rb'})
(1-(-w)^n\frac{\rb'}{\r})}{(1-(-w)^n\frac{\rb}{\r'})
(1-(-w)^n\frac{\r'}{\rb})} \label{KM}
\ee
As in section 2 the log square terms must be understood as
\be
\Big(\ln \frac{\r}{\r'}\Big)^2 = \frac 14 \Big(\ln \big(\frac{\r}{\r'}
\big)^2\Big)^2 \0
\ee
and so on.

Notice that $\G_\M^{ij}(\rho,\rho') = \G_\M^{ji}(\rho',\rho)$.
It is now quite a standard matter to verify that eqs.(\ref{greenrhoB})
and (\ref{bcrhoB}) are satisfied. It is also easy to verify that
the continuity condition on the boundary of the M\"{o}bius band is satisfied:
\be
\G_\M^{ij}(1,\r') = \G_\M^{ij}(-w,\r'), \quad \quad \G_\M^{ij}(-1,\r') =
 \G_\M^{ij}(w,\r'),\quad \quad\forall \, \rho'\0
\ee

It remains for us to discuss Gauss's theorem. The normal derivatives
along the boundary $AA'BB'$ of fig.1 are very complicated. We limit 
ourselves here to reporting the result (see Appendix A for notation):
\be
&& \int_{A'}^B dl\, \d_\perp \G(\r,\r') + \int_{B'}^A dl\, \d_\perp
\G(\r,\r') = 0\label{gaussB}\\
&& \int_A^{A'} dl \, \d_\perp \G(\r,\r')+ 
\int_{B'}^B dl\, \d_\perp\G(\r,\r')=
\int_A^{A'} dl \, \d_\perp f(\r,\r')+ \int_{B'}^B dl\, \d_\perp f(\r,\r')=
2\pi\0
\ee
Therefore Gauss's theorem is verified (see the analogous proof in section
2). We notice however that, as in the case of $B=0$, $\G$ is not continuous
across the junction line $A'B$, which is identified with $B'A$ of fig.1. In
fact, instead of
\be 
\G_\M^{ij}(-w\rb,\r') = \G_\M^{ij}(\r,\r'), \label{Mbc}
\ee 
which would be needed in order to satisfy the M\"{o}bius periodicity
conditions, we only have 
\be
\G_\M^{ij}(-w\r,\r') = \G_\M^{ij}(\r,\r'), \label{MTbc}
\ee 
This means that $\G$ is single--valued on the double covering $\widehat \M$
of $\M$. $\widehat \M$ is obtained by adding to the half annulus $AA'BB'$
of fig.1 its complex conjugate region in the upper half plane and
identifying $e^{i\theta}$ with $w e^{i(\pi+\theta)}$ for $0\leq \theta
\leq 2\pi$. The covering projection is obtained by identifying 
$\r$ and $\rb$. The resulting figure is a torus with a M\"{o}bius 
strip inscribed in it\footnote{This, of course, does not mean that the 
propagator $\G$ is the string propagator on the torus, because of the
boundary conditions (\ref{bcrhoB})}. 

When we restrict our consideration
to the half annulus $AA'BB'$, $\G$ satisfies all the requirements,
including Gauss's theorem, but has a finite discontinuity along the junction
line. We should therefore ask ourselves if this discontinuity may have any
physical consequences. In string theory open string amplitudes depend 
on the propagator on the boundary of $\M$, not on the values taken
by the propagator in the bulk. Now the limit to the boundary of $\M$ is 
well defined and the discontinuity disappears. Therefore the discontinuity
across the junction line does not seem to entail any physical consequence.
On the other hand, if we consider the electrostatic analog of section 2,
we see that the electric field turns out to be discontinuous along the
junction, and, in this case, a physical interpretation is possible only
on the double covering $\widehat \M$.

Since in this paper we are interested in open string amplitudes we will
assume that the right object to be considered is the restriction of $\G$
to the boundary of $\M$.
By taking the 
limit for $\rho$ and $\rho'$ approaching the real axis we get: 
$\G_{\M}^{ij} \rightarrow G_{\M}^{ij}$, where
 \be
G_{\M}^{ij}(\rho,\rho')= 2\alpha'\hat g^{ij} G_\M(\rho,\rho') -\frac i2  
\theta^{ij}
 \epsilon(\rho-\rho')\label{boundpropB}
\ee 
and $G_\M(\rho,\rho')$ is the same as in section 2, 
eqs.(\ref{I+M}, \ref{I-M}).
This is the propagator we will use for our calculations in the 
following section.

Finally we notice that by replacing $(-w)^n$ with $w^n$ in (\ref{GM})
we get the Green function for the annulus, from which one
can extract the planar and nonplanar propagators. This was done in
\cite{ACNY} and in \cite{BCR} and we will rely on those results.

To complete this section we write down the expression of the above 
M\"{o}bius propagator in the $z$ plane. The latter is obtained from
(\ref{GM}, \ref{IM}, \ref{JM}, \ref{fM}, \ref{KM}), passing from $\r$ to $z$, changing
$\tau\to -1/\tau$ and using well--known identities for the Jacobi 
theta--functions, \cite{GSW}: 
\be
\!\!\!\!\frac {1}{\alpha'}\G_\M^{ij}(z,z') = g^{ij}\left(\I_\M(z,z') + 
f_\M(z,z')\right) 
 + (2{\hat g}^{ij}-g^{ij})\J_\M(z,z')   
 + \frac{\theta^{ij}}{\a'}\K_\M(z,z')\label{GMz}
\ee
where 
\be 
\I_\M(z,z') &=& \ln \left\vert\left(\frac{z}{z'}\right)^{\frac 14}-
\left(\frac{z'}{z}\right)^{\frac 14}\right\vert
+ \ln \prod_{n=1}^{\infty}\frac{\bigg\vert 1-(-\sqrt q)^n\sqrt{\frac{z}{z'}}
\bigg\vert   \cdot
\left\vert 1-(-\sqrt q)^n\sqrt{\frac{z'}{z}}\right\vert}{(1-(-\sqrt q)^n)^2}
\0\\
&&\0 \\
f_\M(z,z') &=& -\ln \,\vert zz' \vert \0\\
\J_\M(z,z') &=&
\ln \left\vert\left({z}{\zb'}\right)^{\frac 14}-
\left({\zb'}{z}\right)^{-\frac 14}\right\vert+\ln \prod_{n=1}^{\infty}
\frac{\left\vert1-(-\sqrt q)^n\sqrt{{z}{\zb'}}\right\vert\cdot
\left\vert1-(-\sqrt q)^n\frac 1{\sqrt{{\zb'}{z}}}\right\vert}{(1-(-\sqrt q)^n)^2}
\0\\
&&\0\\
\K_\M(z,z') &=&  
\ln \frac{\left({z}{\zb'}\right)^{\frac 14}-
\left({z}{\zb'}\right)^{-\frac 14}}
{\left({z'}{\zb}\right)^{\frac 14}-
\left({z'}{\zb}\right)^{-\frac 14}}
+ \ln \prod_{n=1}^{\infty}
\frac{(1-(-\sqrt q)^n\sqrt{{z}{\zb'}}) 
(1-(-\sqrt q)^n\frac 1{\sqrt{{\zb'}{z}}})}{(1-(-\sqrt q)^n\sqrt{{\zb}{z'}})
(1-(-\sqrt q)^n\frac 1{\sqrt{{z'}{\zb}}})}
\0\\
&&\0
\ee
where $q= {\rm exp}[- \pi^2/\tau]$. Actually the expression for $\K_\M(z,z')$ 
differs from (\ref{KM}) by a constant term, which is within the 
ambiguity allowed by the Green function's defining equations.
If $F=0$ and we restrict the above expressions to the boundary, i.e.
$|z|=|z'|=1$, $\I_\M$ becomes identical to $\J_\M$ and the propagator reduces (up to
an additive constant) to the expression one can find in \cite{GSW}. 
The expression (\ref{GMz}) of the Green function shows that it is indeed defined on the 
M\"{o}bius band since it  can be thought as the ``bulk counterpart''
of (\ref{expl}). If we express the $z$ coordinate  in terms of $\hat{\nu}$, obtaining $z=\exp{2\pi i \hat{\nu}}$, we see that (\ref{GMz}) has double period with respect to the analogous expression for the Green function on the annulus in presence of a $B$ field, presented for instance in eq. (2.21) of \cite{BCR}.

\section{Field theory limit of gluon amplitudes with $B$ field} 

Switching on a constant $B$ field, on the basis of the discussion in 
previous section, amounts to replacing the propagator used in
section 3 with the full propagator (\ref{boundpropB}). Inserting it into the
general formula
 (\ref{1lampl}) has a simple effect. The addition of the
second term
 $-\frac i2 \theta^{ij}\epsilon(\r-\r')$ does not affect
derivatives of
 propagators, while it modifies the term $\prod_{r<s}
e^{p_rG(\r_r-\r_s)p_s}$.
 This modification turns out to be very simple since
the insertion points
 along the boundary of $\M$ are ordered, so that the
relevant $\epsilon$
 function is always either +1 or --1. As a consequence the
corresponding
 exponential factors can be extracted from the moduli integral. 
In other words, the gluon amplitudes are multiplied by a global 
(\nc) factor
\be
A^{(1)}(p_1,\ldots,p_m)\rightarrow \prod_{r<s} e^{p_r \times p_s}
A^{(1)}(p_1,\ldots,p_m)\label{mf}
\ee
where $A^{(1)}(p_1,\ldots,p_m)$ are the $B=0$ amplitudes and 
$p\times q = \frac i2 p_i \theta^{ij}q_j$. The same is true
also at tree level, \cite{SW,BSST}, and, on the basis of \cite{CRS}, it is 
likely to hold at any loop order, although we do not try to prove it here.

We can now infer that the analysis of the singularities in the
field theory limit does not change with respect to the previous section,
except for the global noncommutative factor in (\ref{mf}). We 
can therefore conclude that the structure of the divergent terms, 
as well as the
renormalization constants, are the same as in the ordinary $SO(N)$ gauge
theories. Therefore, if there exists a \nc gauge field theory that
represents the low energy effective action of open strings with 
orthogonal CP factors in the presence of a constant $B$ field, 
{\it this \nc gauge field theory is one--loop renormalizable}.

\section{Discussion}

The above conclusion seems to imply that a renormalizable \nc gauge field
theory with
 $so(N)$ Chan--Paton factors should exist. We recall that, 
even without resorting to an action,
we can extract the gluon Feynman rules for this low energy field theory from
the string tree amplitudes. They are as follows

\vskip .2cm
\noindent{\bf gluon propagator}. 
\begin{eqnarray}
~~~~\parbox{30mm}{
\begin{fmfchar*}(60,30)
   \fmfleft{i}
   \fmfright{o}
   \fmf{photon, label=$p$, labe.side=left}{i,o}
   \fmflabel{$A,i$}{i}
   \fmflabel{$B,j$}{o}
\end{fmfchar*}}~~~&~~~&~~~ -\frac {i}{p^2}\, \delta_{ab}\, \hat g_{ij}
\end{eqnarray}

\vskip .4cm
\noindent{\bf 3--gluon vertex}. The external gluons carry labels $(a,i,p)$, 
$(b,j,q)$ and $(c,k,r)$ for the Lie algebra, momentum and 
Lorentz indices and are ordered in anticlockwise sense: 
\vspace{5mm}

\begin{center}
\begin{fmfchar*}(85,85)
   \fmftop{i1}
   \fmfbottom{o1,o2}
   \fmf{photon}{i1,v1}
   \fmf{photon}{o1,v1}
   \fmf{photon}{o2,v1}
   \fmflabel{$a,i, p$}{i1}
   \fmflabel{$c, k, r$}{o2}
   \fmflabel{$b, j, q$}{o1}
\end{fmfchar*}
\end{center}
\be
-gf^{abc}\,\cos (p\times q)  
\left(\hat g_{ij}\,(p-q)_k
+\hat g_{jk}\,(q-r)_i+\,\hat g_{ki}(r-p)_j\right)\label{3gluv}
\ee

\vspace{5mm}
\noindent {\bf 4--gluon vertex}.  The gluons carry labels $(a,i,p)$,
$(b,j,q)$, $(c,k,r)$ and $(d,l,s)$ for Lie algebra, Lorentz index 
and momentum. They are clockwise ordered:
\vspace{6mm}

\begin{center}
\begin{fmfchar*}(90,90)
   \fmftop{i1,i2}
   \fmfbottom{o1,o2} 
   \fmf{photon}{i1,v1}
   \fmf{photon}{o1,v1}
   \fmf{photon}{o2,v1}
   \fmf{photon}{i2,v1}
   \fmflabel{$a,i, p$}{i1}
   \fmflabel{$d, l, s$}{o1}
   \fmflabel{$b, j, q$}{i2}
   \fmflabel{$c, k, r$}{o2}
\end{fmfchar*}
\end{center}

\be
-ig^2 &&\bigg\{\bigg[\, \frac{}{}f^{xab}f^{xcd}\cos(p\times q)\cos(r\times s)
\0\\
&&-\left. \left(4d^{abcd}-
\frac 13 (f^{xac}f^{xbd}+f^{xbc} f^{xad})\right)\sin(p\times q)\sin
(r\times s)\right]
\left(\hat g_{ik}\hat g_{jl}-\hat g_{il}\hat g_{jk}\right)\0\\
&&+\bigg[ f^{xac}f^{xdb}\cos(p\times r)\cos(s\times q)
\label{4gluon}\\
&&-\left. \left(4d^{abcd}-
\frac 13 (f^{xcd}f^{xab}+f^{xcb} f^{xad})\right)\sin(p\times r)\sin
(s\times q)\right]
\left(\hat g_{il}\hat g_{jk}-\hat g_{ij}\hat g_{kl}\right)\0\\
&&+\bigg[ f^{xad}f^{xbc}\cos(p\times s)\cos(q\times r)
\0\\
&&-\left.\left. \left(4d^{abcd}-
\frac 13 (f^{xdb}f^{xac}+f^{xba} f^{xdc})\right)\sin(p\times s)\sin
(q\times r)\right]
\left(\hat g_{ij}\hat g_{kl}-\hat g_{ik}\hat g_{jl}\right)\right\}\0
\ee
We recall that this last vertex can be obtained from the string four--gluon
amplitude only after subtracting two suitable tree one--particle reducible 
diagrams. 

One can verify that the above Feynman diagrams can be obtained from the 
action suggested in \cite{BSST}. From that action, which was called
$NCSO(N)$, one can in addition 
extract the Feynman rules for the ghost fields. A natural question
that arises is whether by applying these Feynman rules to compute
one--loop amplitudes one gets the same results as the ones we obtained
in the previous section. The surprising answer is that, if we apply Feynman 
rules in the ordinary way, we get a different result.

To illustrate the problem the simple $NCSO(2)$ case will do. From the string 
theory point of view it is rather easy to argue that the theory should not 
have UV divergences. Let us summarize our previous analysis.
The one--loop contributions to open string amplitudes with $SO(N)$ Chan--Paton
factors are of three types: planar (P) and nonplanar (NP) with the 
world--sheet of the annulus, and nonorientable
(NO) with the world--sheet of the M\"{o}bius strip. Due to the structure
of the string propagators on the annulus and on the M\"{o}bius strip,
the contributions in the presence and in the absence of the B field
for P and NO differ only by overall \nc factors. It follows that those 
contributions which become divergent
in the field theory limit are the same whether B is there or not.
Now in the ordinary $SO(N)$ case the divergent part comes
from the planar contribution with a factor of $N$ in front, and 
from the NO contribution with a factor of $-2$. So altogether the 
divergent field theory part is proportional to $N-2$, and therefore
vanishes in the case $N=2$. This is obvious from the ordinary
field theory side, because the theory is free. However, as we noticed
above, this conclusion holds also in the \nc case. Therefore the
$NCSO(2)$ theory should not give rise to UV divergences. 

Now let us look at the one--loop order on the \nc field theory side.
The Feynman rules are very simple in this case since only the four--point
vertex is nonvanishing. Let us rewrite the four--gluon vertex adapted
to this case
\be
-2ig^2 &&\left[\,\, \cos(p\times r-q\times s)\, (\hat g_{ik}\hat g_{jl}+
\hat g_{ij}\hat g_{kl}-2\hat g_{il}\hat g_{jk})\right.\nonumber\\
&&\! + \,\cos(p\times s+q\times r)\, (\hat g_{il}\hat g_{jk}+
\hat g_{ik}\hat g_{jl}-2\hat g_{ij}\hat g_{kl})\label{4p}\\
&&\! + \left. \cos(p\times s-q\times r)\, (\hat g_{ij}\hat g_{kl}+
\hat g_{il}\hat g_{jk}-2\hat g_{ik}\hat g_{jl})\,\right]\nonumber
\ee
The one--loop correction is infinite. So the theory needs a
renormalization. What is worse is that the divergent part is
not of the form (\ref{4p}), but
\be
\sim \frac{g^4}{\epsilon}&&\left[\,\,\cos(p\times r-q\times s)\, 
(7\hat g_{ik}\hat g_{jl}+
7\hat g_{ij}\hat g_{kl}-8\hat g_{il}\hat g_{jk})\right.\nonumber\\
&&\!+ \,\cos(p\times s+q\times r)\, (7\hat g_{il}\hat g_{jk}+
7\hat g_{ik}\hat g_{jl}-8\hat g_{ij}\hat g_{kl})\label{4p2}\\
&&\!+ \left. \cos(p\times s-q\times r)\, (7\hat g_{ij}\hat g_{kl}+
7\hat g_{il}\hat g_{jk}-8\hat g_{ik}\hat g_{jl})\,\right]\nonumber
\ee
In order to eliminate this divergence we need a counterterm of the
form
\be
\sim (7A_i*A^i*A_j*A^j-4A_i*A_j*A^i*A^j)
\ee
Therefore not only the $NCSO(2)$ gauge field theory is not finite, but 
the divergent part breaks the gauge symmetry. One might argue that
$NCSO(N)$ gauge theories are nonlocal theories and it is perhaps too
much hoping for another miracle like the renormalizability of
\nc $U(N)$ theories to happen also in this case. However the fact
the string theory with $so(N)$ CP factors in the presence of a 
$B$ field is well--behaved and its field theory limit is well--defined,
suggests another possible solution to the puzzle. After a moment's thought
one realizes that the element where field theory and string theory diverge   
is not the Feynman rules themselves (or the action they come from) but 
their application in the one--loop calculation. We have applied them
in the usual way, but that may be too naive. We would need a suitably
modified set of rules. However so far we have
not been able to modify the Feynman rules in such a way as to
reconcile \nc field theory with the results from string theory.
It should be recalled at this point that this reconciliation
is certainly desirable but it might not be possible 
(without violating any fundamental principle, like locality, since
the theory we are dealing with is nonlocal).
If this turns out to actually be the case, it means that we have
found an example of a discrepancy between string theory and the 
corresponding effective (\nc) field theory at one--loop.

\acknowledgments
 
We would like to thank  T.Krajewski, A.Lerda, R.Russo,
S.Sciuto, M.Sheikh--Jabbari  and A.Tomasiello for useful discussions. 
This work was partially supported  by the Italian MURST for the program 
``Fisica Teorica delle Interazioni Fondamentali''.

\appendix

\section{M\"{o}bius strip notation}

In section 2 and 4 we use the following notation for differentiation and
integration on $\M$. We write $\r= x+iy=r e^{i\theta}$. Then
\be
\d_r = \frac 1r (\r \d_\r + \rb \d_{\rb} ),\quad\quad \d_\theta =i(\r\d_\r
- \rb\d_{\rb})\0
\ee
and
\be
\d_x^2+\d_y^2 = 4 \d_\r \d_{\rb}
= \d_r^2 +\frac 1r \d_r +\frac {1}{r^2}
\d_\theta^2\0
\ee
The normal derivatives and line elements along the boundary of fig.1
are defined as follows
\be
 \d_\perp &=& \frac 1r \d_\theta , \quad\quad dl = dr \quad\quad {\rm along}
\quad AA'\0\\
\d_\perp &=& -\frac 1r \d_\theta , \quad\quad dl = -dr \quad\quad {\rm along}
\quad B'B\0\\
\d_\perp &=&  \d_r , \quad\quad dl = -d\theta \quad\quad {\rm along}
\quad A'B\0\\
\d_\perp &=&  -\d_r , \quad\quad dl = w\,d\theta \quad\quad {\rm along}
\quad B'A\0
\ee

\section{ $SO(N)$ tensors}

In this Appendix we collect the conventions relevant for the $so(N)$ Lie 
algebra tensors and traces. We denote the Lie algebra generators by $t^a$,
where $a=1,\dots, \frac {N(N-1)}2$. They are real antisymmetric matrices
with Lie bracket and normalization defined by
\be
[t^a,t^b]= f^{abc}t^c,\quad\quad\quad \tr (t^at^b)= -\frac 12 \delta^{ab}
\label{norm}
\ee
$\tr$ is the trace in the fundamental representation and summation over 
repeated indices is understood. With these conventions we find
\be
\tr(t^at^bt^c) = - \frac 14 f^{abc}\0
\ee
Unlike the $u(N)$ Lie algebra, $so(N)$ does not possess a third order 
invariant symmetric tensor. The fourth order invariant symmetric tensor
is defined by means of
\be
{\rm Sym}(t^at^bt^c) \equiv \frac 16 \left(t^at^bt^c + 
5 \,\,{\rm permutations}\right) \equiv d^{abcd}t^d\label{d4}
\ee
We find
\be
\tr(t^at^bt^ct^d)= -\frac 12 d^{abcd} -\frac 16 f^{abx}f^{xcd}+ 
\frac 1{12} f^{xac}f^{xbd}\label{tr4t}
\ee
Evaluating one--loop Feynman diagrams in field theory requires the
corresponding traces in the adjoint representation. Let us denote  
by $F^a$ the matrices
\be
(F^a)_{bc}= f^{abc}\0
\ee
and by $\Tr$ the traces in the vector space of the adjoint representation.
Then one finds

\be
\Tr (F^aF^b)= -\frac{1}{2} (N-2) \delta^{ab}~, &&
~~~\Tr (F^aF^bF^c)= \frac 14 (N-2) f^{abc}\label{tr3F} \\
\ee
and
\be
\Tr (F^aF^bF^cF^d)&=& -\frac {N-2}{2}d^{abcd} + \frac 14 \left(\delta^{ab}  
\delta^{cd} + \delta^{ac}\delta^{bd}+\delta^{ad}\delta^{bc}\right)\0\\
&&+\frac{N-2}{12} \left(f^{adx}f^{xbc}-f^{abx}f^{xcd}\right)\label{tr4F}
\ee

\end{fmffile}

\begin{thebibliography}{99}

\bibitem{connes} A.Connes, ``Noncommutative Geometry,'' Academic Press 
(1994).

\bibitem{filk} T.Filk, ``Divergencies in a field theory on quantum space,''
{\it Phys.\ Lett.} {\bf B376} (1996) 53.

\bibitem{VGB} J. C. Varilly and J. M. Gracia-Bondia, ``On the ultraviolet 
behaviour of quantum fields over noncommutative manifolds,''
{\it Int.\ J.\ Mod.\ Phys.} {\bf A14} (1999) 1305, hep-th/9804001.

\bibitem{CDP} M. Chaichian, A. Demichev and P. Presnajder, ``Quantum Field 
Theory on Noncommutative Space-Times and the Persistence of Ultraviolet 
Divergences,'' {\it Nucl.\ Phys.} {\bf B567} (2000) 360, hep-th/9812180.

\bibitem{MSR} C.P. Martin and D. Sanchez-Ruiz, ``The One-loop UV Divergent 
Structure of U(1) Yang-Mills Theory on Noncommutative $R^4$,'' 
{\it Phys.\ Rev.\ Lett.} {\bf 83} (1999) 476-479, hep-th/9903077.

\bibitem{jabbari} M.Sheikh--Jabbari, ``One loop renormalizability of 
supersymmetric 
Yang--Mills theories on \nc two--torus'', {\it JHEP} {\bf 9906} (1999) 015, 
hep-th/9903107.

\bibitem{tomas} T.Krajewski and R.Wulkenhaar, ``Perturbative quantum gauge 
fields on the \nc torus'', 
{\it Int.\ J.\ Mod.\ Phys.} {\bf A15} (2000) 1011, 
hep-th/9903187.
Harald Grosse, Thomas Krajewski and Raimar Wulkenhaar,  
``Renormalization of noncommutative Yang-Mills theories: A simple example'', 
hep-th/0001182.

\bibitem{ABK} I. Ya. Aref'eva, D. M. Belov and A. S.Koshelev, ``A Note on 
UV/IR for Noncommutative Complex Scalar Field,'' hep-th/0001215.
\,\,I. Ya. Aref'eva, D. M. Belov, A. S. Koshelev and O. A. Rytchkov,
`` UV/IR Mixing for Noncommutative Complex Scalar Field Theory, II
(Interaction with Gauge Fields),'' hep-th/0003176. 

\bibitem{haya} M. Hayakawa, ``Perturbative analysis on infrared and  
ultraviolet aspects of noncommutative QED on $R^4$,'' hep-th/9912167. 

\bibitem{CHMS} S. Cho, R. Hinterding, J. Madore and  H. Steinacker,  
``Finite Field Theory on Noncommutative Geometries,''
{\it Int.\ J.\ Mod.\ Phys.} {\bf D9} (2000) 161-199, hep-th/9903239.

\bibitem{CR} I. Chepelev and R. Roiban, ``Convergence Theorem for  
Non-commutative Feynman Graphs and Renormalization,'' 
{\it JHEP} {\bf 0103} (2001) 001, 
hep-th/0008090.  

\bibitem{adi} 
A. Armoni,   
``Comments on Perturbative Dynamics of Non-Commutative Yang-Mills Theory,''  
{\it Nucl.\ Phys. }  {\bf B593} (2001) 229,  
hep-th/0005208. 

\bibitem{GGRS} H. O. Girotti, M. Gomes, V. O. Rivelles and A. J. da Silva,  
``A Consistent Noncommutative Field Theory: the Wess-Zumino Model,''  
{\it Nucl.\ Phys.} {\bf B587} (2000) 299,  
hep-th/0005272. 
 
\bibitem{BoSa} L. Bonora and M. Salizzoni, ``Renormalization of \nc $U(N)$
gauge theories", {\it Phys.\ Lett.} {\bf B504} (2001) 80, hep-th/0011088.

\bibitem{martin} C.P. Martin, D. Sanchez-Ruiz, ``The BRS invariance of 
noncommutative U(N) Yang-Mills theory at the one-loop level", 
{\it Nucl.Phys.} {\bf B598} (2001) 348-370, hep-th/0012024.

\bibitem{micu} A.Micu and M. M. Sheikh-Jabbari, ``Noncommutative $\Phi^4$ 
theory at two loops,''
{\it JHEP} {\bf 0001} (2001) 025, 
 hep-th/0008057. 

\bibitem{Das}
A.~Das and M.~M.~Sheikh-Jabbari,
``Absence of higher order corrections to noncommutative Chern-Simons  coupling,'' {\it JHEP} {\bf 0106} (2001) 028, hep-th/0103139.

\bibitem{Andreev}
O.~Andreev and H.~Dorn,  
``Diagrams of noncommutative $\Phi^3$ theory from string theory,''     
{\it Nucl.\ Phys. }  {\bf B583} (2000) 145,        
hep-th/0003113. 

\bibitem{Kiem}               
Y.~Kiem and S.~Lee,     
``UV/IR mixing in noncommutative field theory via open string loops,'' 
{\it Nucl.\ Phys. }  {\bf B586} (2000) 303,     
hep-th/0003145.    

\bibitem{BCR} A.Bilal, C.-S. Chu and R.Russo, ``String Theory and 
Noncommutative Field Theories at One Loop,'' {\it Nucl.\ Phys. } 
{\bf B582} (2000) 65, hep-th/0003180. 

\bibitem{GKMRS} J. Gomis, M.Kleban, T.Mehen, M.Rangamani and S.Shenker,
``Noncommutative Gauge Dynamics From The String Worldsheet,'' 
{\it JHEP} {\bf 0008} (2000) 011, hep-th/0003215.

\bibitem{CRS} C.-S. Chu, R.Russo and S.Sciuto, ``Multiloop String Amplitudes 
with  B-Field and Noncommutative QFT,'' {\it Nucl.\ Phys. } {\bf B585} (2000) 193,  
hep-th/0004183. 

\bibitem{Kiem2}
Y.~Kiem, S.~Lee and J.~Park,
``Noncommutative field theory from string theory: Two-loop analysis,''
{\it Nucl.\ Phys. }  {\bf B594} (2001) 169, 
[hep-th/0008002].

\bibitem{Arcioni}
G.~Arcioni, J.~L.~Barbon, J.~Gomis and M.~A.~Vazquez-Mozo,
``On the stringy nature of winding modes in noncommutative thermal field  theories,'' {\it JHEP} {\bf 0006} (2000) 038, hep-th/0004080.

\bibitem{MRS} S.~Minwalla, M.~Van Raamsdonk and N.~Seiberg, ``Noncommutative 
perturbative dynamics,'' 
{\it JHEP} {\bf 0002} (2000) 020, 
hep-th/9912072. 
M.~Van Raamsdonk and N.~Seiberg, ``Comments on \nc perturbative dynamics,'' 
{\it JHEP} {\bf 0003} (2000) 035, hep-th/0002186.

\bibitem{MST} A.Matusis, L.Susskind and N.Toumbas, ``The IR/UV Connection in 
the Non-Commutative Gauge Theories,''
{\it JHEP} {\bf 0012} (2000) 002,  
hep-th/0002075.

\bibitem{trek} H.Liu and J.Michelson, ``*--Trek: The one--loop ${\cal N}=4$
\nc SYM action,''  
{\it Nucl.\ Phys.\ }  {\bf B614} (2001) 279,  hep-th/0008205.

\bibitem{zanon} D.Zanon, ``Noncommutative perturbation in superspace,''
{\it Phys.\ Lett. } {\bf B504} (2001) 101, 
hep-th/0009196; 
A.Santambrogio and D.Zanon, ``One--loop four--point function
in \nc ${\cal N}=4$ Yang--Mills theory,'' 
{\it JHEP} {\bf 0101} (2001) 024, 
hep-th/0010275.
M.~Pernici, A.~Santambrogio and D.~Zanon, 
``The one-loop effective action of noncommutative N = 4 super Yang-Mills  
is gauge invariant,''
{\it Phys.\ Lett. }  {\bf B504} (2001) 131, 
hep-th/0011140.
D.~Zanon,
``Noncommutative N = 1,2 super U(N) Yang-Mills: UV/IR mixing and  effective action results at one loop,''
{\it Phys.\ Lett.}  {\bf B502} (2001) 265,  
hep-th/0012009.

\bibitem{griguolo} L. Griguolo and M. Pietroni, ``Hard \nc loop resummation",
hep-th/0102070; 
L. Griguolo and M. Pietroni,
``Wilsonian renormalization group and the non-commutative IR/UV  connection,'' 
{\it JHEP} {\bf 05} (2001) 032, hep-th/0104217; 
A. Bassetto, L. Griguolo, G.Nardelli and F.Vian, ``On the unitarity of quantum
gauge theories in \nc spaces", {\it JHEP} {\bf 0107} (2001) 008, hep-th/0105257.

\bibitem{ChuKhoTrav} 
C.~Chu, V.~V.~Khoze and G.~Travaglini, 
``Dynamical Breaking of Supersymmetry in Noncommutative Gauge Theories,'' 
{\it  Phys.\ Lett. }  {\bf B513} (2001) 200, 
hep-th/0105187. 
 
\bibitem{J&J} 
I.~Jack and D.~R.~Jones, ``Ultra-violet Finite Noncommutative Theories,''  
hep-th/0105221. 

\bibitem{harvey} 
N. A. Nekrasov, ``Trieste lectures on solitons in \nc gauge theories",
hep-th/0011095. 
A. Konechny and A. Schwarz, ``Introduction to M(atrix) theory and \nc
geometry", hep-th/0012145. 
J. H. Harvey, ``Komaba lectures on \nc solitons and 
D--branes", hep-th/0102076. 

\bibitem{witten} E. Witten, ``Noncommutative Tachyons And String Field 
Theory",  hep-th/0006071.
M. Schnabl, ``String field theory at large B-field and noncommutative 
geometry", 
{\it JHEP} {\bf 0011} (2000) 031, hep-th/0010034.

\bibitem{BSST} L. Bonora, M. Schnabl, M.M. Sheikh--Jabbari and A. Tomasiello, 
``Noncommutative SO(n) and Sp(n) gauge theories,'' {\it Nucl.\ Phys. } {\bf B589} (2000) 461, hep-th/0006191. 

\bibitem{BarsShahin}
I.~Bars, M.~M.~Sheikh-Jabbari and M.~A.~Vasiliev,
``Noncommutative O*(N) and usp*(2N) algebras and the corresponding gauge  field theories,'' {\it Phys.\ Rev.\ }  {\bf D64} (2001) 086004, 
hep-th/0103209.

\bibitem{MSSW} J.Madore, S.Schraml, P.Schupp and J.Wess, ``Gauge theory 
on noncommutative spaces,'' 
{\it Eur.\ Phys.\ J. }  {\bf C16} (2000) 161, hep-th/0001203.

\bibitem{BS} 
M.Bianchi and A.Sagnotti,  
``On The Systematics Of Open String Theories,''
{\it Phys.\ Lett.} {\bf B247} (1990) 517; 
M.Bianchi and A.Sagnotti, 
``Twist symmetry and open string Wilson lines,''
{\it Nucl. Phys.} {\bf B361} (1991) 519.

\bibitem{SW} N. Seiberg and E. Witten, ``String theory and \nc geometry",
{\it JHEP} {\bf 09} (1999) 032, hep-th/9908142. 

\bibitem{botu} R.Bott and L.Tu,`` Differential forms in algebraic
topology,'' Springer Verlag Inc., New York, (1982).

\bibitem{ACNY} A.Abouelsaood, C.G.Callan, C.R.Nappi and S.A.Yost,  
``Open strings in background gauge fields,'' {\it Nucl.\ Phys.} {\bf B280} 
(1987) 599. 

\bibitem{GSW} 
M.B. Green, J.H. Schwarz and E. Witten, ``Superstring theory,'' 
Cambridge, Uk: Univ. Pr. (1987). 

\bibitem{VMLRM} P. Di Vecchia, L. Magnea, A.Lerda, R.Russo and 
R.Marotta, ``String techniques for the calculation of renormalization
constant in field theory,'' {\it Nucl.\ Phys. }  {\bf B469} (1996) 235,
hep-th/9601143.


\bibitem{abbot} L.F. Abbot, ``The background field method beyond one loop,''
{\it Nucl.\ Phys. } {\bf B185} (1981) 189.

\bibitem{FT}
E.~S.~Fradkin and A.~A.~Tseytlin,
``Nonlinear Electrodynamics From Quantized Strings,''
{\it Phys.\ Lett. }  {\bf B163} (1985) 123.


\bibitem{GNSS} D.J. Gross, A. Neveu, J.Scherk and J.H. Schwarz, 
``Renormalization 
and unitarity in the dual resonance model,''
 {\it Phys.\ Rev. } {\bf D2} (1970) 697.



\end{thebibliography}
\end{document}